\documentclass{aa}  
\usepackage{graphicx}
\usepackage{subfig}
\usepackage{txfonts}
\usepackage{natbib}

\begin{document}

   \title{Asteroseismic modelling of the two F-type hybrid pulsators KIC\,10080943A and KIC\,10080943B}

   \author{V.~S. Schmid\inst{1}\fnmsep\thanks{Aspirant PhD Fellow of the Research Foundation Flanders (FWO), Belgium}
          \and
          C. Aerts\inst{1,2}
          }

\institute{
   Institute for Astronomy, KU Leuven, Celestijnenlaan 200D, B -- 3001 Leuven, Belgium\\
   \email{valentina.schmid@ster.kuleuven.be}
   \and
   Department of Astrophysics/IMAPP, Radboud University Nijmegen, P.O. Box 9010, 6500 GL Nijmegen, The Netherlands
             }

   \date{Received; accepted}

 
  \abstract
   {Pulsating binary stars are ideal targets for testing the theory of stellar structure and evolution. Fundamental parameters can be derived to high precision from binary
modelling and provide crucial constraints for seismic modelling. High-order gravity modes are sensitive to the conditions near the convective core and therefore allow for a determination of parameters describing interior physics, especially the convective-core overshooting parameter. KIC\,10080943 is a binary system that contains two gravity- and pressure-mode hybrid pulsators. A detailed observational study has provided fundamental and seismic parameters for both components.}
   {We aim to find a model that is able to predict the observed g-mode period spacings and stellar parameters of both components of KIC\,10080943.}
   {By calculating model grids with the stellar evolution code MESA and the seismic code GYRE, we can compare theoretical properties to the observed mean period spacing and position in the Hertzsprung-Russell diagram.}
   {The masses of our best models are somewhat below the values estimated from binarity, which is a consequence of the low observed mean g-mode period spacing. We find that the amount of core overshooting and diffusive mixing can be well constrained by the equal-age requirement for the two stars, however, we find no significant difference for different shapes of core overshooting. The measured rotation rates are within the limit of validity for the first-order perturbation approximation. We can find a good fit by using the traditional approximation for the pulsations, when taking slightly younger models with a higher asymptotic period spacing. This is because the zonal modes experience a slight shift due to the Coriolis force, which the first-order perturbation approximation ignores. }
   {}


   \titlerunning{Asteroseismic modelling of KIC\,10080943A and KIC\,10080943B}
   \authorrunning{V.\,S. Schmid \& C. Aerts}
   
   \maketitle
%

\section{Introduction}

Stellar models play an important role in aiding astrophysical research in a wide variety of contexts, despite several shortcomings and uncertainties in the input physics. Fortunately, we can test and improve these models by comparing observed stellar pulsations, that is the asteroseismic fingerprints of the stellar interior, to theoretical predictions. It is imperative to choose the test cases wisely by the amount of observational constraints they hold. As such, gravity (g) and pressure (p) mode hybrid pulsators have a high potential for testing theories of stellar structure and evolution.

The periods of high-order g~modes of a non-rotating, unevolved star with a convective core and a radiative envelope are predicted to be equally spaced \citep{Tassoul1980}. Yet, deviations from this equidistant spacing are expected as the star evolves or rotates \citep{Miglio2008,Bouabid2013}. The extent of the convective core and chemical composition gradients at the boundary of that core influence the mean value and structure of the period spacing, while rotation introduces a tilt in the period spacing pattern. This is of particular interest for stars of intermediate mass ($1.5\leq M~(M_\odot)\leq2.2$) that inhabit the transition region, where a radiative central region gives way to a convective core and the convective envelope becomes increasingly thinner. For these types of stars the g~modes are thought to be excited by the convective flux blocking mechanism \citep{Guzik2000,Dupret2005}. The heat-driven p~modes of hybrid pulsators can then provide an additional selection criterion based on the predictions of mode excitation. Very recently, there have been several studies exploiting the seismic potential of g-mode pulsators with carefully identified modes from four years of \textit{Kepler} \citep{Borucki2010} data \citep[e.g.][]{Kurtz2014,Moravveji2015,Papics2015,Saio2015,VanReeth2015a,VanReeth2015b,Moravveji2016}.

The main caveat of these studies is that they only concern single stars. Their derived fundamental parameters and ages are therefore still dependent on the accuracy of stellar models on which the seismic properties are based. An additional and independent test can be provided by pulsating binary stars, whose measurements of fundamental parameters rely only on Kepler's laws and geometry. Double-lined spectroscopic binaries, where radial velocities of both components are measured, yield a mass ratio, while eclipsing binaries allow for the determination of absolute masses and radii. Such systems are rare and require a dedicated, detailed study to recover the seismic properties of the pulsating star and precise stellar parameters. There have only been a handful of published analyses so far \citep[e.g.][]{Maceroni2009,Maceroni2014,Welsh2011,Hambleton2013}.

In this paper we present the seismic modelling of two hybrid pulsators; KIC\,10080943A and KIC\,10080943B. They reside in a double-lined spectroscopic binary system and have very similar seismic properties and stellar parameters, which are described in Sect.~\ref{sec:observations} and whose analysis has been published by \citet{Schmid2015}. We use the one-dimensional (1D) stellar evolution code MESA \citep{Paxton2011,Paxton2013,Paxton2015} and the stellar pulsation code GYRE \citep{Townsend2013} to find the best representation of both stars, assuming two coeval single stars and neglecting their tidal influence. The modelling process is explained in detail in Sect.~\ref{sec:modelling}. Finally, we present a discussion of our results in Sect.~\ref{sec:discussion}.

\section{Observational constraints on KIC\,10080943}
\label{sec:observations}

\begin{figure*}[t!]
\centering
\includegraphics{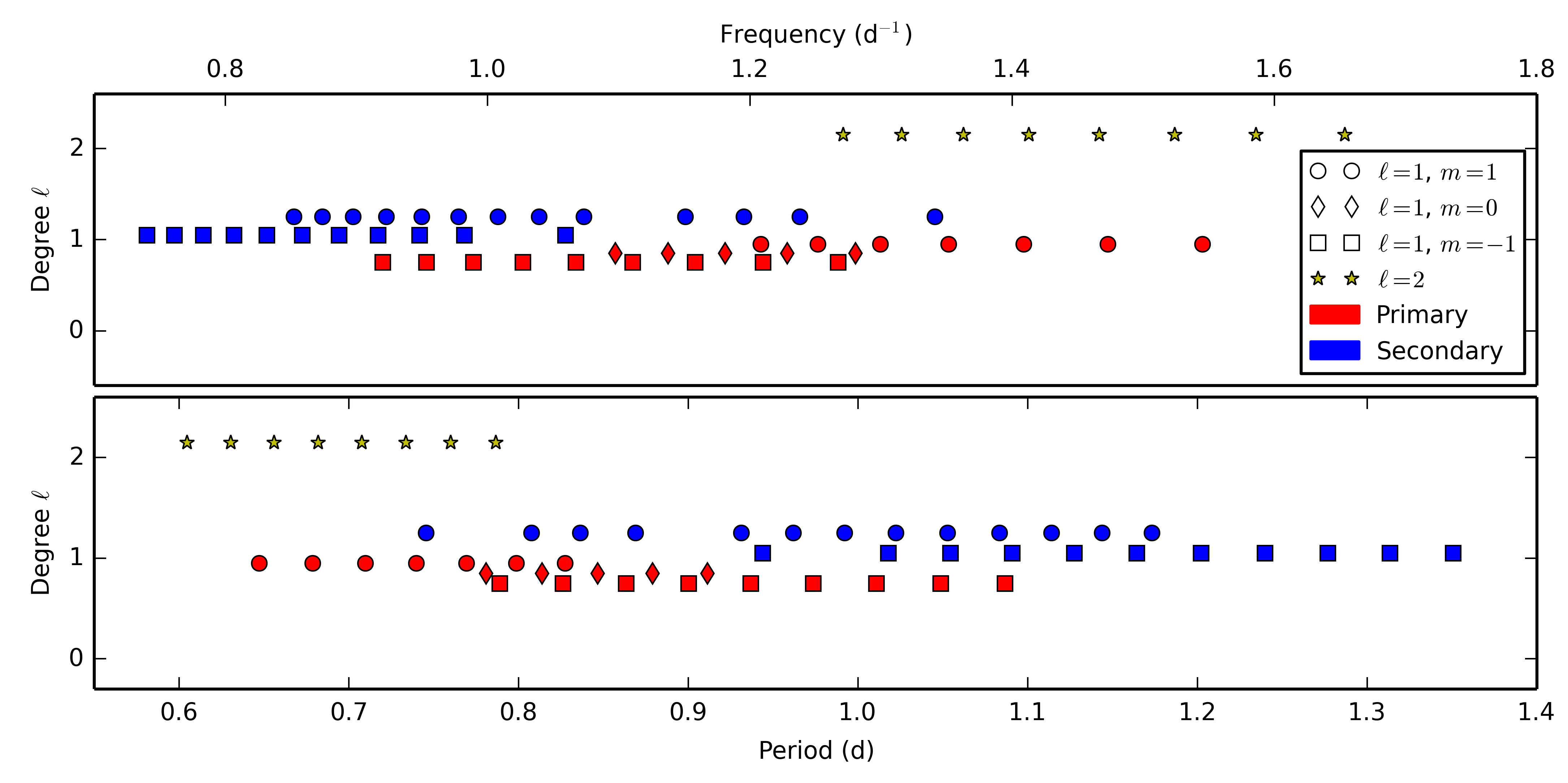}
\caption{Observed g~modes in frequency (top) and period (bottom) as a function of degree $\ell$ are shown for the primary in red and the secondary in blue. The $m$ values are distinguished by the different symbols. The yellow stars denote the $\ell=2$ modes, which could not be assigned to either component.}
\label{fig:gmodes_observed}
\end{figure*}

\begin{figure*}[t!]
\centering
\includegraphics{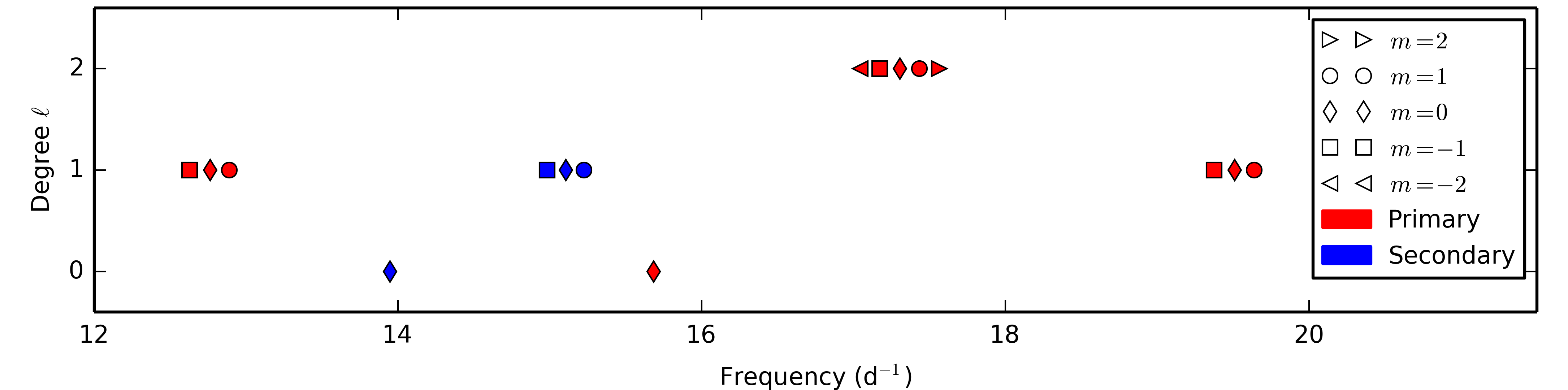}
\caption{Observed p-mode frequencies as a function of degree $\ell$, which was estimated, assuming that all observed multiplets are complete. Modes of the primary are shown in red and those of the secondary are shown in blue. The $m$ values are distinguished by the different symbols.}
\label{fig:pmodes_observed}
\end{figure*}

KIC\,10080943 is an eccentric, double-lined binary system, containing two $\gamma$\,Dor/$\delta$\,Sct hybrid pulsators \citep[for a detailed description of $\gamma$\,Dor and $\delta$\,Sct type pulsators, see][chapt.~2]{Aerts2010}. \citet{Keen2015} published the analysis of the period spacings of the g~modes from \textit{Kepler} photometry. \citet{Schmid2015} performed a detailed analysis of the system; these authors derived fundamental parameters of both components from modelling the binary signals of ellipsoidal variation and reflection in the \textit{Kepler} light curve in addition to the radial velocity curves and disentangled component spectra from high-resolution follow-up spectroscopy covering the orbit. They were able to assign most period spacing series and high-amplitude p~modes to either the primary or the secondary. This was achieved by analysing the phase modulations of the p~modes that arise from the binary motion. Modes of the primary have a time delay of $36.2\pm4.8$~s, while those of the secondary are delayed by $44.0\pm3.5$~s. Combination frequencies between the p~modes and two period spacing series, which form a series of rotationally split doublets, helped to determine that these belong to the secondary. Three other period spacing series form series of triplets with a different rotational splitting value and thus originate in the primary. Figure~\ref{fig:gmodes_observed} shows the six period spacing series and which star they originate in. The series of $\ell=2$ modes could not be assigned to either star, as it does not show combination frequencies or rotational splitting. In Fig.~\ref{fig:pmodes_observed} the p~modes that can be connected to one star or the other are shown.

All derived stellar parameters including the mean period spacing values are listed in Table~\ref{tab:observations} and are used to place both stars in a Hertzsprung-Russell diagram (HRD) inside a 1-$\sigma$ error box in Fig.~\ref{fig:HRDbest}. It is obvious that the parameters lack the precision typical for double-lined eclipsing binaries \citep{Southworth2012}, as no eclipses can be observed for KIC\,10080943. The radii are then derived from the equipotential surfaces in Roche geometry, as opposed to the width of the eclipses. The ratio of the effective temperatures is somewhat constrained by both effects, while they are largely independent of the orbital inclination \citep[for more information on the binary and a description of the modelling process, see][]{Schmid2015}. This high uncertainty of the orbital inclination propagates into the precision of the absolute masses. The mass ratio of the two components, however, was derived with great confidence from the ratio of the radial velocity amplitudes. Hence, the binarity of the system still yields crucial additional constraints for the theoretical stellar models compared to single star pulsators, as is discussed below.

\begin{table}
\caption{Fundamental and seismic parameters of the primary and secondary component of KIC\,10080943.}
\label{tab:observations}
\centering
\begin{tabular}{l c c}
\hline\hline 
 & Primary & Secondary \\ 
\hline
$q=M_2/M_1$ & \multicolumn{2}{c}{$0.96\pm0.01$} \\
$M$ ($M_\odot$) & $2.0\pm0.1$ & $1.9\pm0.1$ \\
$R$ ($R_\odot$) & $2.9\pm0.1$ & $2.1\pm0.2$ \\
$T_\mathrm{eff}$ (K) & $7100\pm200$ & $7480^{+180}_{-200}$ \\
$\log g$ (cgs) & $3.81\pm0.03$ & $4.1\pm0.1$ \\
$\mathrm{[M/H]}$ & -0.05$\pm$0.17 & -0.09$\pm$0.30 \\
$\log (L/L_\odot)$ & $1.28\pm0.06$ & $1.1\pm0.1$ \\
$\Delta P$ (s) & $2817\pm26$ & $2905\pm36$ \\
\hline
\end{tabular}
\end{table}

The main diagnostic for finding a model representative of KIC\,10080943A and KIC\,10080943B is the period spacing of the high-order g~modes. For both components, the mean period spacing value is rather low at $\Delta P<3000$~seconds for stars near $2~M_\odot$. Indeed, comparing with the sample of $\gamma$\,Dor stars in \citet{VanReeth2015b}, the measured $\Delta P$ values of both components hint at masses $<2~M_\odot$. Furthermore, the observed period spacing series show a significant wavy structure without clear dips and a slight tilt, which is caused by rotation. This points towards evolved stars with a low central hydrogen mass fraction, where the receding convective hydrogen-burning core has left behind a chemical composition gradient $\nabla_\mu$ leading to a peak in the Brunt-V\"ais\"al\"a frequency $N$ and therefore a decrease in the asymptotic period spacing \citep[e.g.][]{Miglio2008}.

In addition to the g~modes, two p~modes were identified as candidate radial modes and can help distinguish between possible models. To estimate their radial overtone, we calculate the pulsation constant $Q$ \citep{Breger1990} as
\begin{equation}
Q=P\sqrt{\frac{\bar{\rho}}{\bar{\rho_\odot}}},
\end{equation}
or, expressed in observable parameters
\begin{equation}
\log Q=-6.454+\log P+0.5\log g + 0.1M_\mathrm{bol}+\log T_\mathrm{eff}.
\end{equation}
For the highest amplitude p mode ($f_{s,2}=13.947586~\mathrm{d}^{-1}$) and for the parameters of the secondary, we find $Q=0.0332\pm0.0045$, which is indicative of the radial fundamental mode. The p mode with frequency $f_{s,1}=15.68333~\mathrm{d}^{-1}$ belongs to the primary and has $Q=0.018\pm0.001$, which is consistent with $3\leq n\leq5$ \citep{Stellingwerf1979}.

In addition to these seismic diagnostics, the main constraint from the binarity is the requirement of equal age, equal composition, and the mass ratio. This means that the seismic properties of both models have to match the observations at the same age, while the ratio of their masses also has to be in agreement with the observational measurement.

\section{Searching for the best seismic models}
\label{sec:modelling}

We used MESA \citep[version 7385]{Paxton2011,Paxton2013,Paxton2015} to calculate evolutionary tracks for different parameters, which are specified in the following subsection. For all tracks we used OPAL opacity tables \citep{Iglesias1993,Iglesias1996} constructed for the solar metal mixture provided by \citet{Asplund2009}. Stellar layers are dynamically unstable against convection according to the Schwartzschild criterion and are modelled with the mixing length theory (MLT) by \citet[chap.~14]{Cox1968}, using $\alpha_\mathrm{MLT}=1.8$. All equilibrium models along the evolutionary tracks are non-rotating, as both components are slow rotators \citep[rotational period near the core $P_{c,1}\approx 7$ d and $P_{c,2}\approx 11$ d,][]{Keen2015}. Our MESA inlist is provided in Appendix~\ref{app:inlist} and is available for further choices of input physics.

As already mentioned in Sect.~\ref{sec:observations}, the Brunt-V\"ais\"al\"a frequency increases with the steepness of the chemical composition gradient $\nabla_\mu$. The shape and strength of the mixing of the stellar material therefore influence the shape of $N$ and hence the periods and period spacings of high-order g~modes. This can be illustrated by the asymptotic period spacing, which is defined by \citet{Tassoul1980} as
\begin{equation}
\Delta P_c=\frac{2\pi^2}{L\int^1_{x_0}\frac{|N|}{x}\mathrm{d}x},
\label{eq:dp_asym}
\end{equation}
where $L=\sqrt{\ell(\ell+1)}$, $x=r/R$ and $x_0$ is the convective-core boundary. To probe the strength of mixing beyond the convective regions, we included  convective-core overshooting and diffusive mixing as free parameters into our models. Convective-core overshooting mainly increases the size of the convective core, thus decreasing the integral $\int|N|/x\mathrm{d}x$. It is implemented in MESA using two different definitions that both depend on the MLT diffusion coefficient ($D_\mathrm{conv}$) and an adjustable parameter. In the exponential description, the region immediately outside the core has a diffusion coefficient as determined by
\begin{equation}
D_\mathrm{ov}=D_\mathrm{conv}\exp\left(-\frac{2z}{f_\mathrm{ov}H_{P}}\right),
\end{equation}
where $H_{P}$ is the local pressure scale height, $z$ is the distance from the convective-core boundary, and $f_\mathrm{ov}$ is a free parameter \citep{Herwig2000}. This definition results in an exponentially decaying $D_\mathrm{ov}$. Alternatively overshooting can also be defined as a step function, where the extent of the overshooting layer is $d_\mathrm{ov}=\alpha_\mathrm{ov}H_{P}$ with the adjustable parameter $\alpha_\mathrm{ov}$. Typically $\alpha_\mathrm{ov}\simeq10f_\mathrm{ov}$ and increasing either parameter leads to a wider mixed region beyond the core. In the radiative parts of the model, where neither convection nor overshooting is active, we set a minimum diffusion coefficient $D_\mathrm{mix}$ of $1~\mathrm{cm}^2\,\mathrm{s}^{-1}$. It is implemented such that the diffusion coefficient throughout the star does not fall below this value. The higher it is, the more the chemical composition gradient is washed out in the radial direction, leading to fewer dips in the period spacing pattern. Apart from affecting $\nabla_\mu$, both overshooting and diffusion also have the consequence of mixing more hydrogen into the central region and therefore extending the time the stars spend on the main sequence. In other words, otherwise equal stars have a higher $X_c$ at a given age when only the mixing strength is increased.

\begin{table}[t!]
\caption{Extents of the grid parameters.}
\label{tab:grid}
\centering
\begin{tabular}{l l l l r}
\hline\hline
 & begin & end & step & \multicolumn{1}{c}{$N$} \\
\hline
\multicolumn{4}{l}{\textit{coarse grid}} & 704 \\
$M$ ($M_\odot$) & 1.7 & 2.2 & 0.05 & 11 \\
$Z$ & 0.012 & 0.018 & 0.002 & 4 \\
$f_\mathrm{ov}$ & 0.0 & 0.03 & 0.01 & 4 \\
$\log D_\mathrm{mix}$ & 0 & 3 & 1 & 4 \\
 & & & \\
\multicolumn{4}{l}{\textit{fine grid (exponential overshooting)}} & 3267 \\
$M$ ($M_\odot$) & 1.76 & 1.86 & 0.01 & 11 \\
$Z$ & 0.0105 & 0.0125 & 0.001 & 3 \\
$f_\mathrm{ov}$ & 0.005 & 0.015 & 0.001 & 11 \\
$\log D_\mathrm{mix}$ & 0 & 2 & 0.25 & 9 \\
 & & & \\
\multicolumn{4}{l}{\textit{fine grid (step overshooting)}} & 3267 \\
$M$ ($M_\odot$) & 1.76 & 1.86 & 0.01 & 11 \\
$Z$ & 0.0105 & 0.0125 & 0.001 & 3 \\
$\alpha_\mathrm{ov}$ & 0.05 & 0.15 & 0.01 & 11 \\
$\log D_\mathrm{mix}$ & 0 & 2 & 0.25 & 9 \\
\hline
\end{tabular}
\tablefoot{The overshooting parameters $f_\mathrm{ov}$ and $\alpha_\mathrm{ov}$ are expressed in terms of local pressure scale height ($H_P$) and the diffusion coefficient $D_\mathrm{mix}$ has the unit cm$^2$\,s$^{-1}$.}
\end{table}

Based on these stellar models, we calculate theoretical pulsation modes, using the stellar pulsation code GYRE \citep[version 4.3]{Townsend2013}. To take the effect rotation has on the observed frequencies into account, we perturbed the frequencies of the calculated zonal modes. In a first step we apply the first-order perturbative approach $\Delta f=m\beta_{n,\ell}\Omega$ \citep{Ledoux1951}, where $\beta_{n,\ell}=1-C_{n,\ell}$ and $C_{n,\ell}$ is the Ledoux constant. We use $\beta_{n,\ell}$ as calculated by GYRE and the observed frequency splitting $\Delta f$ as a first guess for the rotational frequency $\Omega$. The validity of the perturbative approach for $\gamma$\,Dor stars was tested by \citet{Ballot2010} for different rotation rates. For both stars and the observational estimates of the rotation rates \citep{Schmid2015} $0.14< 2\Omega_c/f_\mathrm{co}<0.27$, where $\Omega_c$ is the rotation rate near the core and $f_\mathrm{co}$ are the g-mode frequencies in the co-rotating frame of reference. This means that they are in the validity range of the perturbative approach to first order.

An alternative method to treat the coupling between rotation and pulsation is the traditional approximation of rotation \citep[TAR; e.g.][]{Townsend2003a}, which is also implemented in GYRE. This method is based on neglecting the horizontal component of the rotation vector and is valid when the horizontal displacement is larger than the radial displacement, which is the case for g~modes in $\gamma$\,Dor stars. \citet{Bouabid2013} studied the influence of the Coriolis force on high-order g~modes, using the TAR, and compared their computations to predictions by the first-order perturbative approach. These authors showed that, even for slow rotation rates, the Coriolis force has an influence on the mode periods, which is stronger at higher radial orders. This force causes a departure from linearity in rotational splitting and also effects the zonal modes, which are unperturbed in the perturbative approach.

\subsection{MESA model grids}
\label{sec:mesa_grids}

\begin{figure}[t!]
\centering
\includegraphics{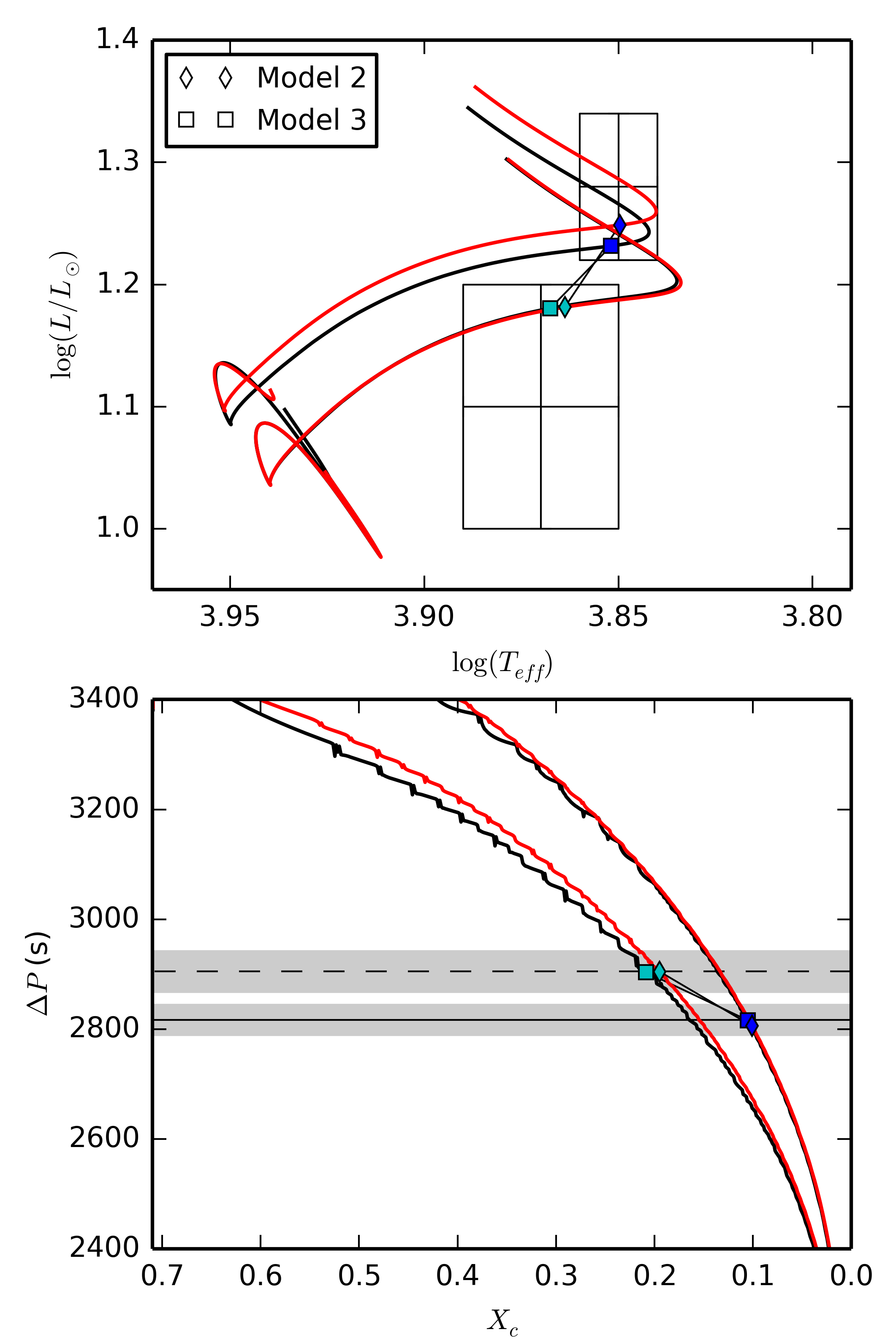}
\caption{\textit{Top panel:} HRD with the observed positions and $1\sigma$ areas of both components of KIC\,10080943, marked by the boxes (the primary is the cooler, more luminous star). The positions of the components in Model 2 are given by the diamonds and for Model 3 by the squares, while the respective evolutionary tracks are given by the red and black solid lines. Dark blue symbols mark the primary, while the light blue symbols indicate the  secondary models. Parameters of these models are given in Table~\ref{tab:bestmodels}. \textit{Lower panel:} Asymptotic period spacing $\Delta P$ as a function of central hydrogen mass fraction $X_c$. The observed mean period spacing of the primary (secondary) is given as the horizontal solid (dashed) line and the $1\sigma$ areas are shown as the grey patch.}
\label{fig:HRDbest}
\end{figure}

\begin{table*}
\caption{Parameters of the best combination of primary and secondary models for each grid of Table~\ref{tab:grid}.}
\label{tab:bestmodels}
\centering
\begin{tabular}{l | r r r r r r}
\hline\hline
 & \multicolumn{1}{c}{Model 1} & \multicolumn{1}{c}{Model 2} & \multicolumn{1}{c}{Model 3} & \multicolumn{1}{c}{Model 4} & \multicolumn{1}{c}{Model 5} & \multicolumn{1}{c}{Model 6} \\
\hline
grid & \multicolumn{1}{c}{\textit{coarse}} & \multicolumn{1}{c}{\textit{exp. ov. (Ledoux)}} & \multicolumn{1}{c}{\textit{step ov. (Ledoux)}} & \multicolumn{1}{c}{\textit{exp. ov. (TAR)}} & \multicolumn{1}{c}{\textit{step ov. (TAR)}} & \multicolumn{1}{c}{\textit{morphology}} \\
$M_1$ ($M_\odot$) & 1.85 & 1.82 & 1.81 & 1.82 & 1.81 & 1.67 \\
$M_2$ ($M_\odot$) & 1.75 & 1.76 & 1.76 & 1.76 & 1.76 & 1.60 \\
$Z$ & 0.012 & 0.0125 & 0.0125 & 0.0125 & 0.0125 & 0.010 \\
$f_\mathrm{ov,1}$ or $\alpha_\mathrm{ov,1}$ & 0.01 & 0.008 & 0.11 & 0.008 & 0.11 & 0.007 \\
$f_\mathrm{ov,2}$ or $\alpha_\mathrm{ov,2}$ & 0.0 & 0.005 & 0.05 & 0.005 & 0.05 & 0.006 \\
$\log(D_\mathrm{mix,1})$ & 0 & 0.25 & 0.25 & 0.25 & 0.25 & 0.5 \\
$\log(D_\mathrm{mix,2})$ & 2 & 1.5 & 1.75 & 1.5 & 1.75 & 0.75 \\
\hline
$a_1$ (Gyr) & 1.087 & 1.123 & 1.110 & 1.102 & 1.089 & 1.082 \\
$a_2$ (Gyr) & 1.082 & 1.127 & 1.110 & 1.100 & 1.081 & 1.085 \\
$X_{c,1}$ & 0.09 & 0.1 & 0.11 & 0.12 & 0.13 & 0.22 \\
$X_{c,2}$ & 0.24 & 0.19 & 0.21& 0.22 & 0.23 & 0.29 \\
$\Delta P_{c,1}$ (s) & 2832 & 2806 & 2816 & 2873 & 2877 & 2829 \\
$\Delta P_{c,2}$ (s) & 2909 & 2905 & 2904 & 2939 & 2942 & 2851 \\
$T_\mathrm{eff,1}$ (K) & 7131 & 7074 & 7110 & 7157 & 7189 & 7583 \\
$T_\mathrm{eff,2}$ (K) & 7526 & 7308 & 7372 & 7387 & 7452 & 7597 \\
$R_1$ ($R_\odot$) & 2.92 & 2.81 & 2.72 & 2.74 & 2.66 & 2.12 \\
$R_2$ ($R_\odot$) & 2.29 & 2.43 & 2.39 & 2.38 & 2.33 & 1.90 \\
$\log (L_1/L_\odot)$ & 1.30 & 1.25 & 1.23 & 1.25 & 1.23 & 1.13 \\
$\log (L_2/L_\odot)$ & 1.18 & 1.18 & 1.18 & 1.18 & 1.18 & 1.04 \\
\hline
\end{tabular}
\tablefoot{The overshooting parameters $f_\mathrm{ov}$ and $\alpha_\mathrm{ov}$ are expressed in terms of local pressure scale height ($H_P$) and the diffusion coefficient $D_\mathrm{mix}$ has the unit cm$^2$\,s$^{-1}$.}
\end{table*}

From the results of the binary modelling by \citet{Schmid2015}, we can draw an error box on the HRD, within which our best models should reside. We calculated a grid of models over the following parameters that influence the position of the model on the HRD: mass $M$, metallicity $Z$, convective-core overshooting $f_\mathrm{ov}$ in the exponential description, and diffusive mixing $D_\mathrm{mix}$ (the extent of the grid is given in Table~\ref{tab:grid}), while the initial hydrogen mass fraction was fixed at the solar value $X_i=0.71$.

At every time step of each track, we calculated the asymptotic period spacing $\Delta P_c$ (Eq.~\ref{eq:dp_asym}). This parameter, as well as the fundamental model parameters $\log (L/L_\odot)$, $\log (R/R_\odot)$, $\log T_\mathrm{eff}$, and $X_c$ were then interpolated on an equidistant age scale with steps of $\Delta a=10^7$ years. At every age step we evaluated models for the primary and the secondary simultaneously by comparing the observed to the asymptotic period spacing, allowing only combinations with equal composition at the zero-age main sequence (ZAMS; $Z_1=Z_2$) and which fulfil the observed mass ratio $M_2/M_1=q=0.96\pm0.01$ within $2\sigma$. We calculated the $\chi^2_\mathrm{tot}$ as
\begin{equation}
\chi^2_\mathrm{tot}=\chi^2_1+\chi^2_2=\left(\frac{\Delta P_c-\Delta P_{o,1}}{\sigma(\Delta P_{o,1})}\right)^2+\left(\frac{\Delta P_c-\Delta P_{o,2}}{\sigma(\Delta P_{o,2})}\right)^2,
\end{equation}
where $\Delta P_{o,1}$ and $\Delta P_{o,2}$ refer to the mean of the observed period spacing of the zonal dipole modes for the primary and the secondary, respectively. For about $4.1\%$ of all 19\,456 valid combinations $\chi^2_1<4$ and $\chi^2_2<4$ at the same age, meaning that the asymptotic period spacing of the model is within $2\sigma$ of the observations. When we compare the model tracks to the observations it becomes obvious that those with high diffusion ($\log D_\mathrm{mix}=3$) are too luminous and too old. Thus, we can limit the number of good models even further, when only considering those whose $\log T_\mathrm{eff}$ and $\log (R/R_\odot)$ lie within $2\sigma$ of the observed values, leaving 60 combinations of primary and secondary models fulfilling all observational constraints derived from the g-mode period spacings and the binarity of our target.

We find that the remaining 60 models have common characteristics. They have low masses (the highest primary mass $M_1=1.9~M_\odot$; the best models have $M_1=1.8-1.85~M_\odot$ and $M_2=1.7-1.75~M_\odot$) and low metallicity, with the majority having $Z=0.012$. While the core overshooting is limited to small values $f_\mathrm{ov}=0.0-0.1$, diffusion with values $\log D_\mathrm{mix}=1-2$ are possible. As already predicted by the mean value and structure of the observed period spacing, both stars are old with $a\approx 1$~Gyr and low central hydrogen mass fractions $X_{c,1}=0.05-0.15$ and $X_{c,2}=0.15-0.3$. The parameters of the best model are given in Table~\ref{tab:bestmodels}.

\begin{figure*}
\centering
\subfloat[Fine grid -- exponential overshoot]{\includegraphics{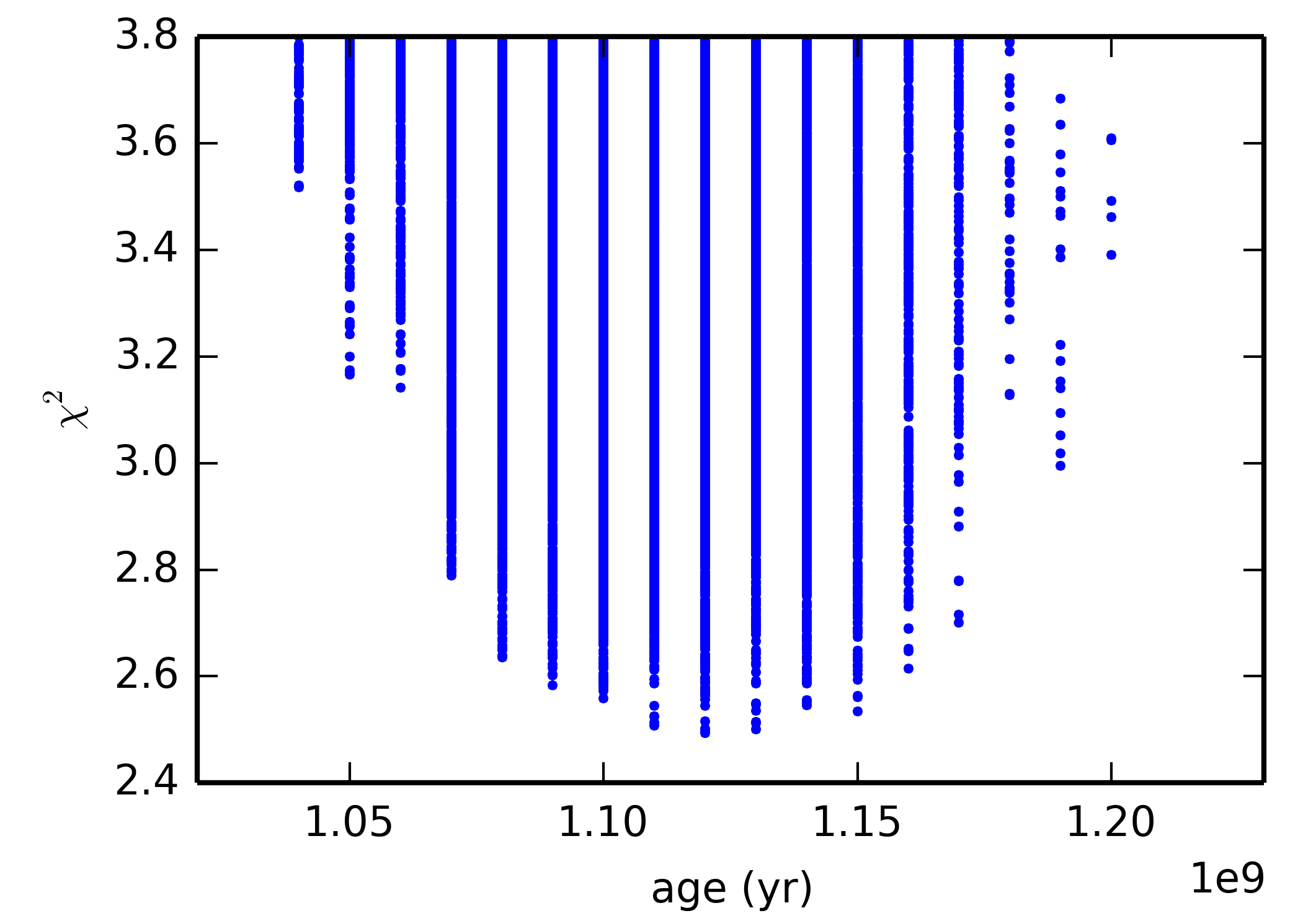}}
\subfloat[Fine grid -- step overshoot]{\includegraphics{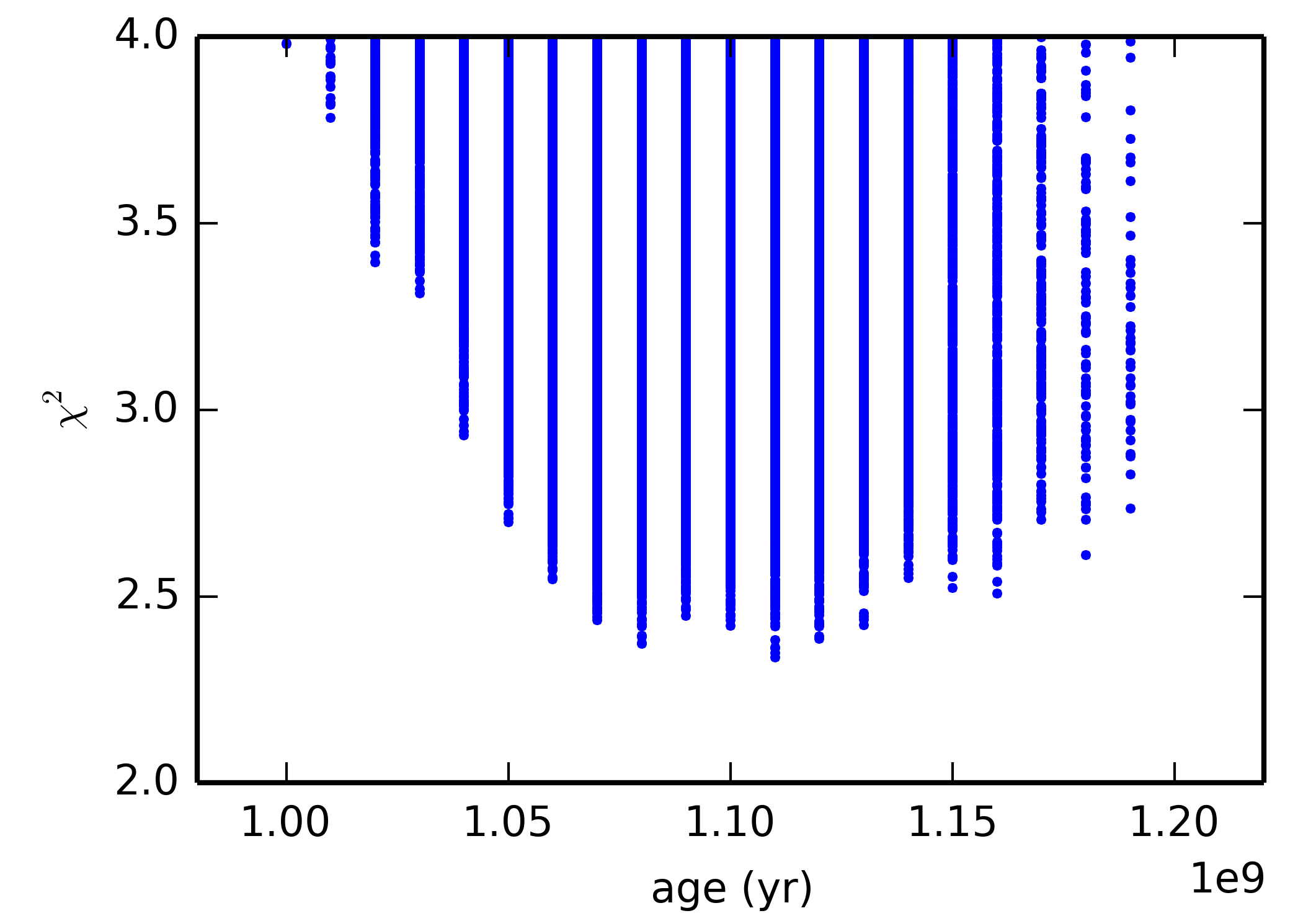}}
\caption{Chi-squared distribution of age for all coeval models in the two denser grids, which fall within $2\sigma$ of the HRD position of both stars and have equal composition.}
\label{fig:chi2age}
\end{figure*}

\begin{figure*}
\centering
\subfloat[Fine grid -- exponential overshoot]{\includegraphics{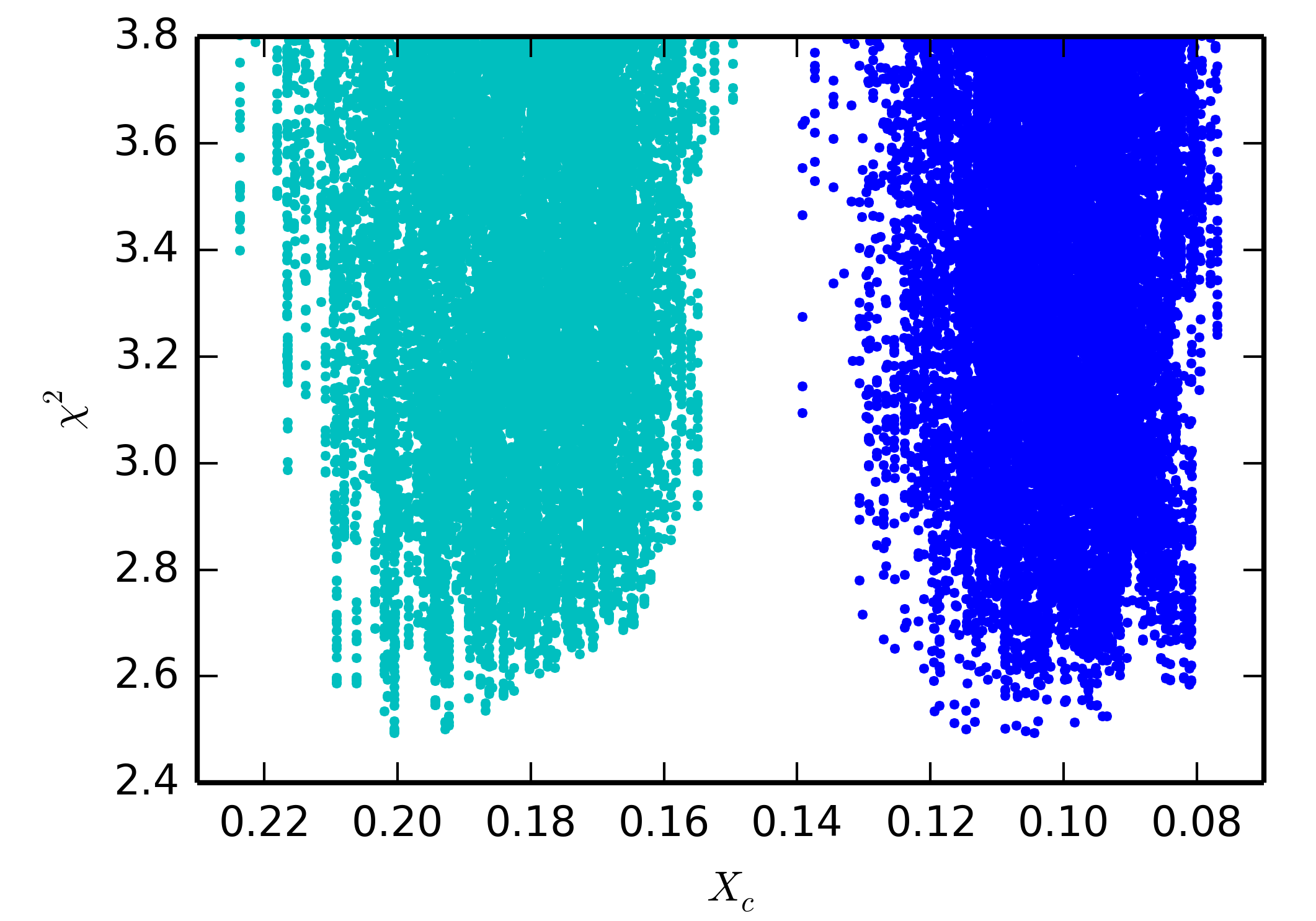}}
\subfloat[Fine grid -- step overshoot]{\includegraphics{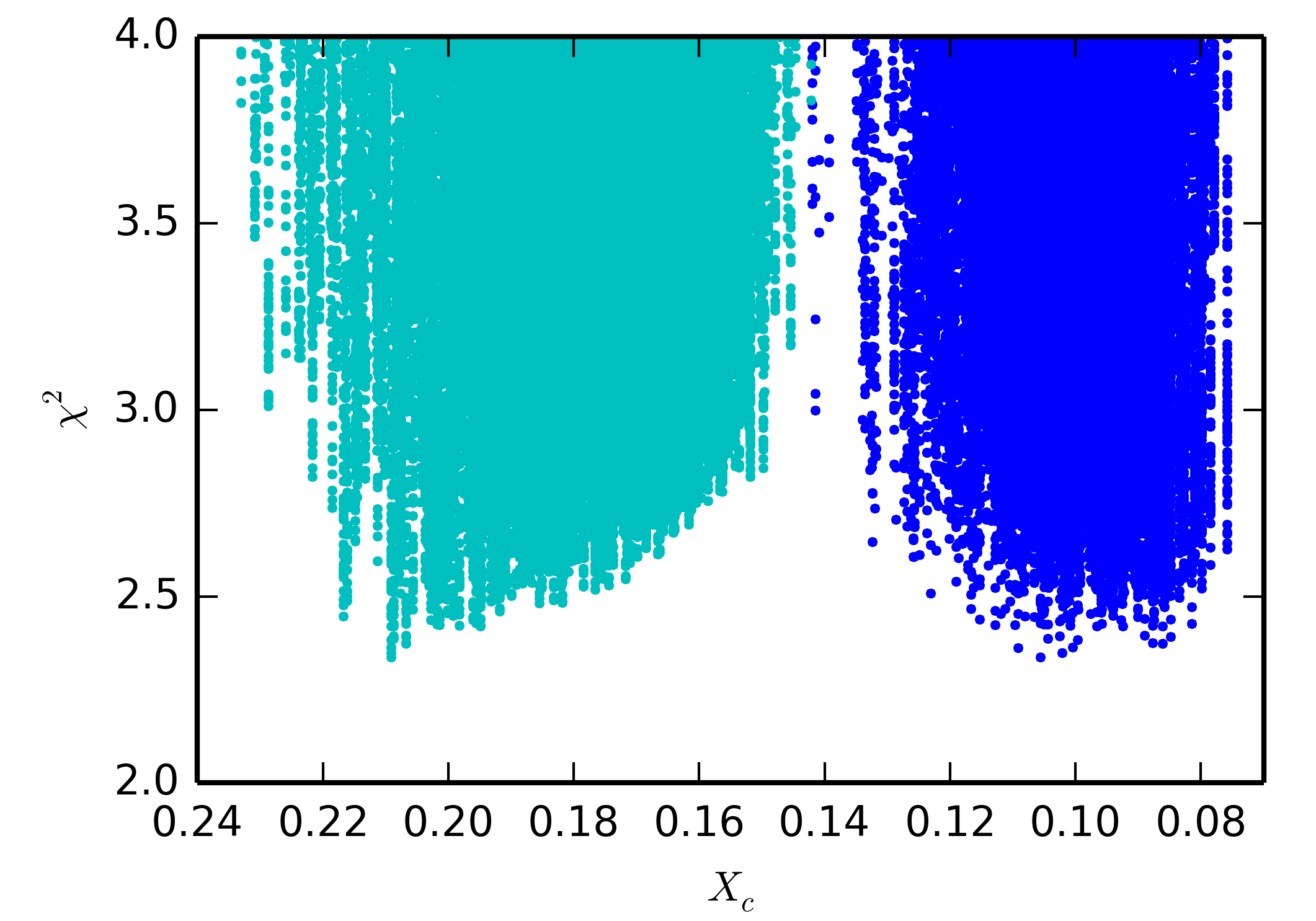}}
\caption{Chi-squared distribution of central hydrogen mass fraction $X_c$ for the same models as in Fig.~\ref{fig:chi2age}. The models for the primary are indicated by the dark blue points, while the models for the secondary are indicated by the light blue points.}
\label{fig:chi2xc}
\end{figure*}

Based on these common characteristics, we computed two denser grids with refined steps in mass and the mixing parameters (see Table~\ref{tab:grid}). The difference between those grids is the definition of core overshooting, where one grid uses the exponential description and the other uses step overshooting. We analysed them using the same method as described above and found 97\,880 and 170\,449 models fulfilling the requirements for the exponential-overshoot and step-overshoot grids, respectively. Their parameters agree well with the results of the coarse grid analysis and can be compared in Table~\ref{tab:bestmodels} for the models of lowest $\chi^2$. Figure~\ref{fig:HRDbest} shows their position and evolutionary tracks in the HRD. In accordance with results from the previous grid, we find that higher values of overshooting combined with low values in envelope diffusion are favoured for the primary, while we find the opposite for the secondary (see Sect.~\ref{sec:discussion} for a discussion on the relevance of these values). Furthermore, we again find that, based on this $\chi^2$ approach, the best models are old at around 1.1~Gyr and have a low $X_c$ (see Figs~\ref{fig:chi2age} and \ref{fig:chi2xc}). 

So far, we have only taken into account the mean of the observed g-mode period spacing, neglecting the effect of rotation and the information that is held in the spacing morphology. In the next step, we therefore calculate the adiabatic frequencies of our models with GYRE and compare their seismic properties to the observations.

\subsection{Period spacing with GYRE}

\begin{figure*}
\centering
\subfloat[Gravity modes of Model 2.]{\label{fig:gmodes_model_lowchi2:a}\includegraphics{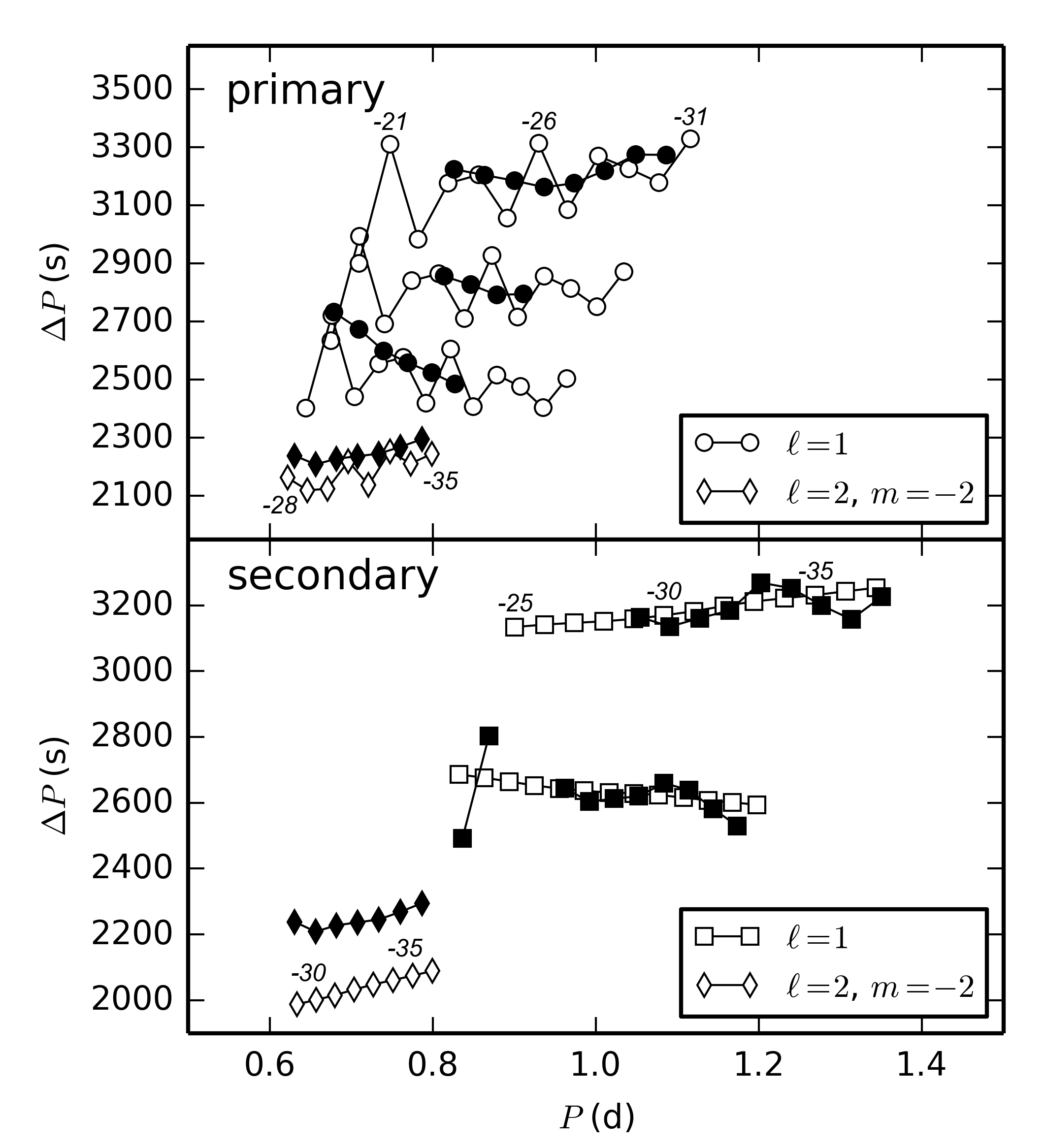}}
\subfloat[Gravity modes of Model 3.]{\includegraphics{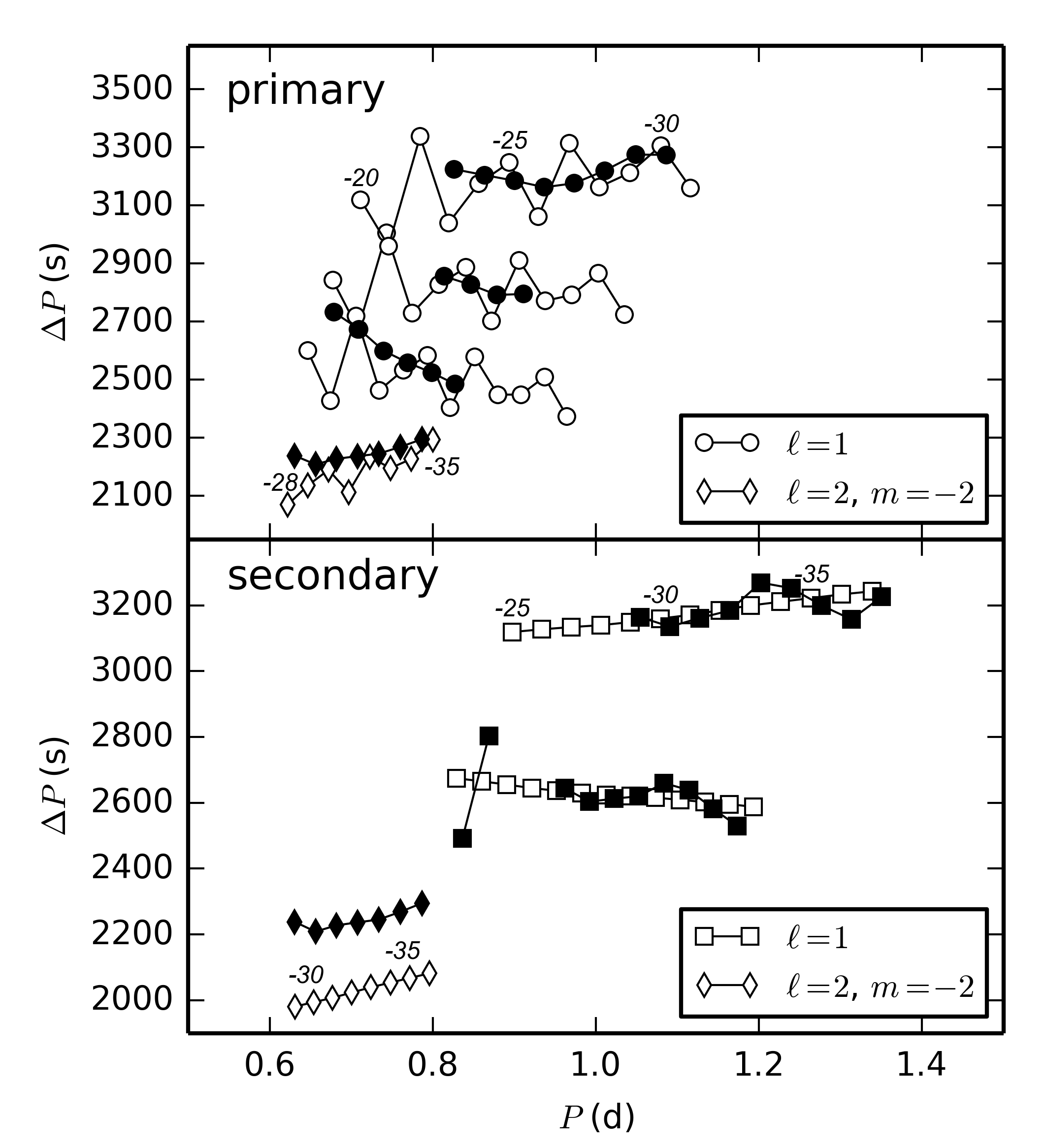}}
\caption{Comparison of theoretical (empty symbols) to observed (filled symbols) period spacing for the primary (top panels) and the secondary (bottom panels). Ledoux splitting has been applied to the theoretical zonal modes to fit the retrograde and prograde modes. The uncertainties of the observed period spacings are smaller than the symbols. The radial orders $k$ are indicated near selected modes. Dipole modes of radial order $-31\leq k \leq -20$ are shown for the primary and $-37\leq k \leq -25$ for the secondary, while the radial orders of the quadrupole modes are $-28\leq k \leq -35$ and $-29\leq k \leq -36$, respectively.}
\label{fig:gmodes_model_lowchi2}
\end{figure*}

\begin{figure*}
\centering
\subfloat[Pressure modes of Model 2.]{\label{fig:pmodes_model:a}\includegraphics{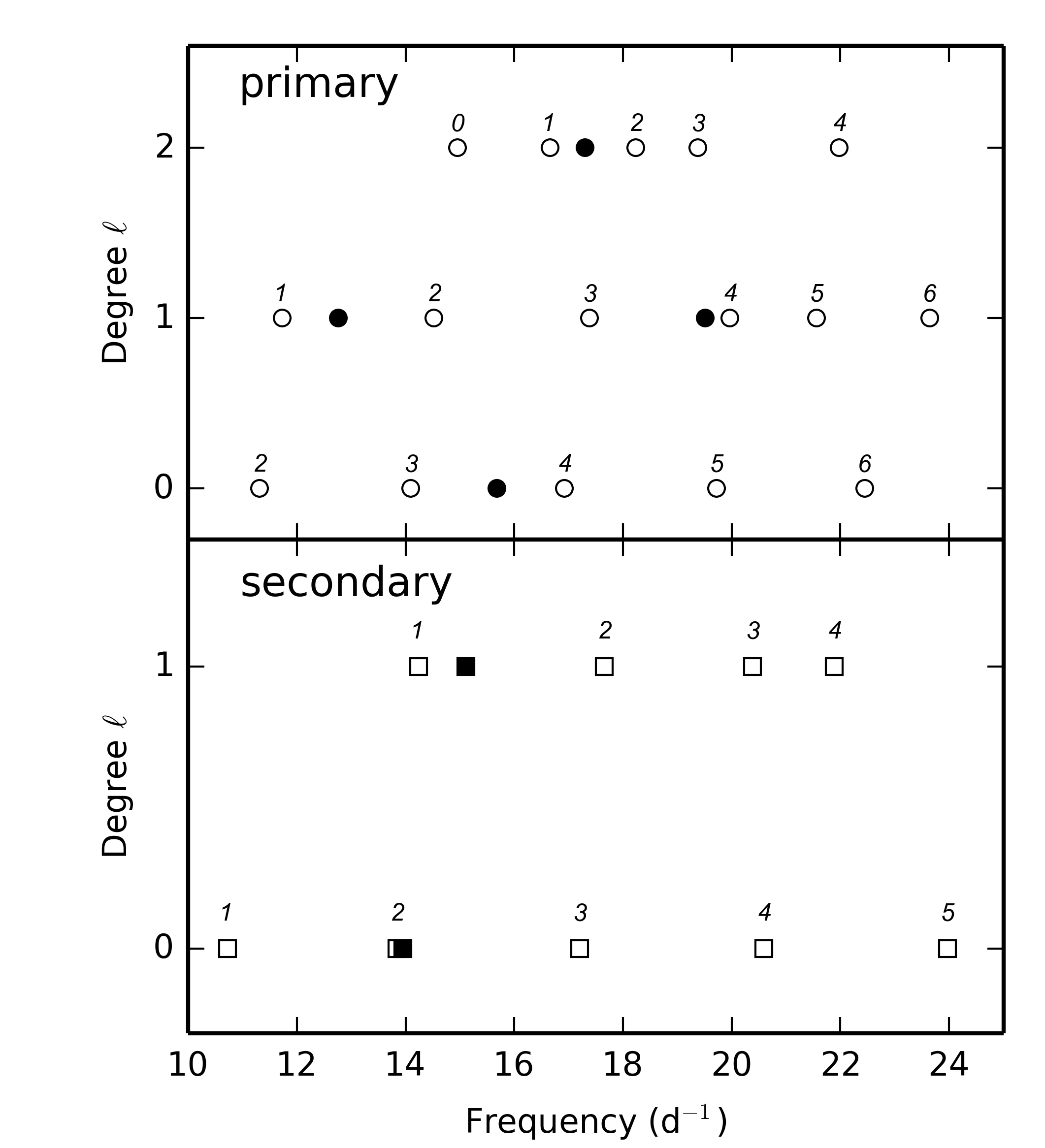}}
\subfloat[Pressure modes of Model 3.]{\label{fig:pmodes_model:b}\includegraphics{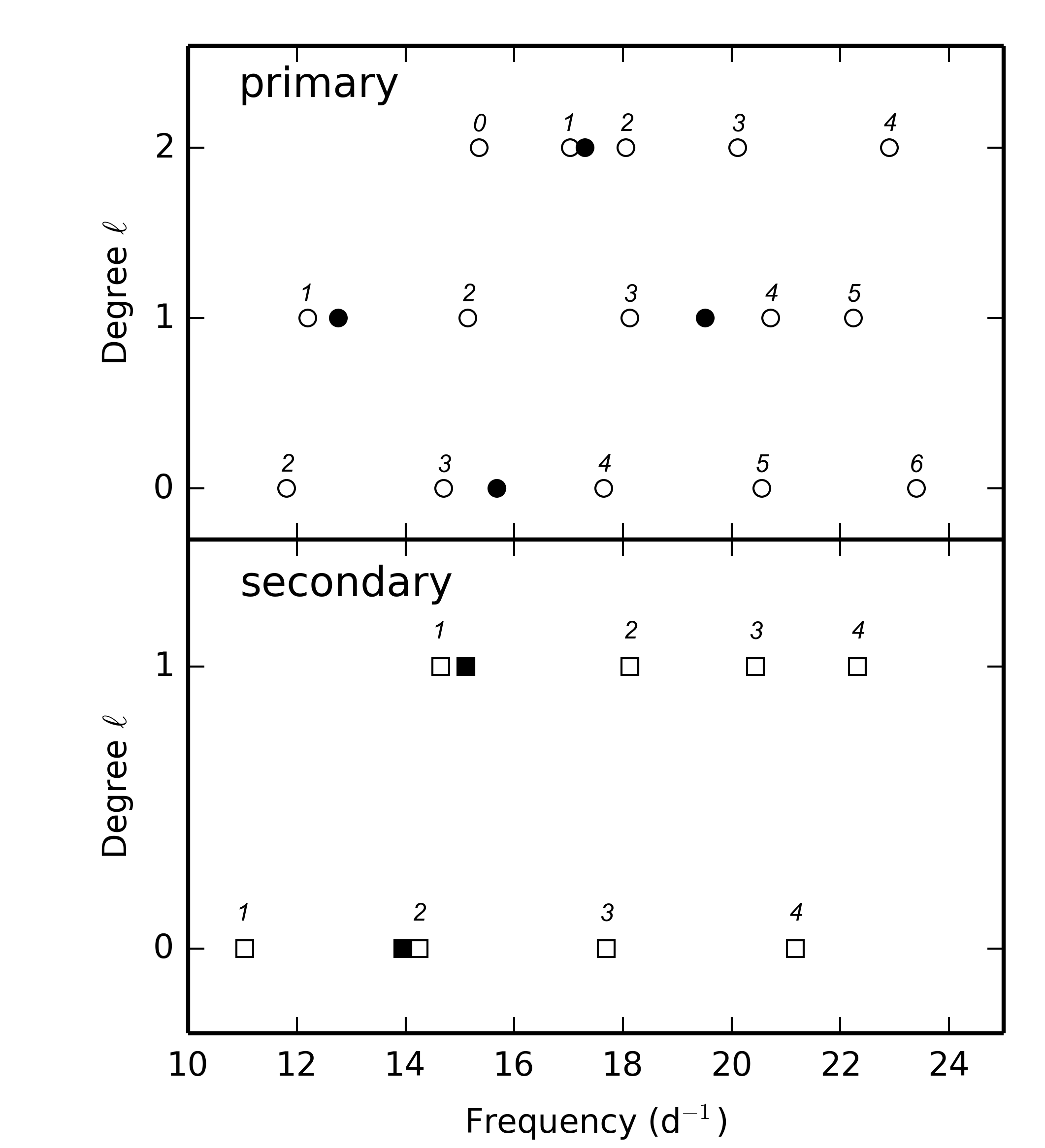}}
\caption{Comparison of the theoretical zonal p~modes (empty symbols) to the observed modes (filled symbols) as a function of degree $\ell$. Modes of the primary are shown in the top panels and for the secondary in the bottom panels. The radial orders are given above the respective modes.}
\label{fig:pmodes_model}
\end{figure*}

We calculated the zonal dipole and quadrupole modes with GYRE for the coeval models with the lowest $\chi^2$. As we clearly see the effect of rotation in the observed frequency spectrum (see Fig.~\ref{fig:gmodes_observed}), we apply Ledoux splitting to the zonal modes to obtain the theoretical prograde and retrograde modes and the rotational frequencies $\Omega_c$ near the core for both components. We derive $\Omega_{c,1}=0.13963\pm0.00007~\mathrm{d}^{-1}$ and $\Omega_{c,2}=0.09034\pm0.00002~\mathrm{d}^{-1}$ and a negligible difference between the models with exponential and those with step overshooting. For these rotation rates and the co-rotating dipole-mode frequencies we find $0.198<2\Omega_1/f_\mathrm{co}<0.264$ and $0.144<2\Omega_2/f_\mathrm{co}<0.237$. The comparison between the calculated modes of Models 2 and 3 and the observed period spacings is illustrated in Fig.~\ref{fig:gmodes_model_lowchi2}, while Fig.~\ref{fig:pmodes_model} shows the comparison of the p~modes. Our main focus lies in comparing the mean period spacing values and the structure of the period spacing pattern, rather than individual frequencies. We find a satisfactory fit for both models and cannot distinguish between exponential or step overshooting. Furthermore, we are able to assign the $\ell=2$ series (yellow stars in Fig.~\ref{fig:gmodes_observed}) to the primary, as there is a clear overlap with the retrograde sectoral modes of both models, while there is a gap for the models of the secondary. From Eq.~(\ref{eq:dp_asym}), we expect that the ratio of the asymptotic period spacing of the quadrupole modes to the asymptotic period spacing of the dipole modes is $\Delta P_c(\ell=2)/\Delta P_c(\ell=1)=0.577$. We use the derived $\Omega_{c,1}$ to shift the observed quadrupole sectoral modes to their zonal counterparts, and find $\Delta P_o(\ell=2)=1664\pm21$~s and $\Delta P_o(\ell=2)/\Delta P_{o,1}(\ell=1)=0.591\pm0.009$. For the secondary and $\Omega_{c,2}$ we find $\Delta P_o(\ell=2)=1840\pm14$~s and $\Delta P_o(\ell=2)/\Delta P_{o,2}(\ell=1)=0.633\pm0.009$. These results support that the $\ell=2$ series originates in the primary.

\begin{figure*}
\centering
\subfloat[Period spacing for Model 2 and Model 4.]{\label{fig:gmodes_ledoux_tar:a}\includegraphics{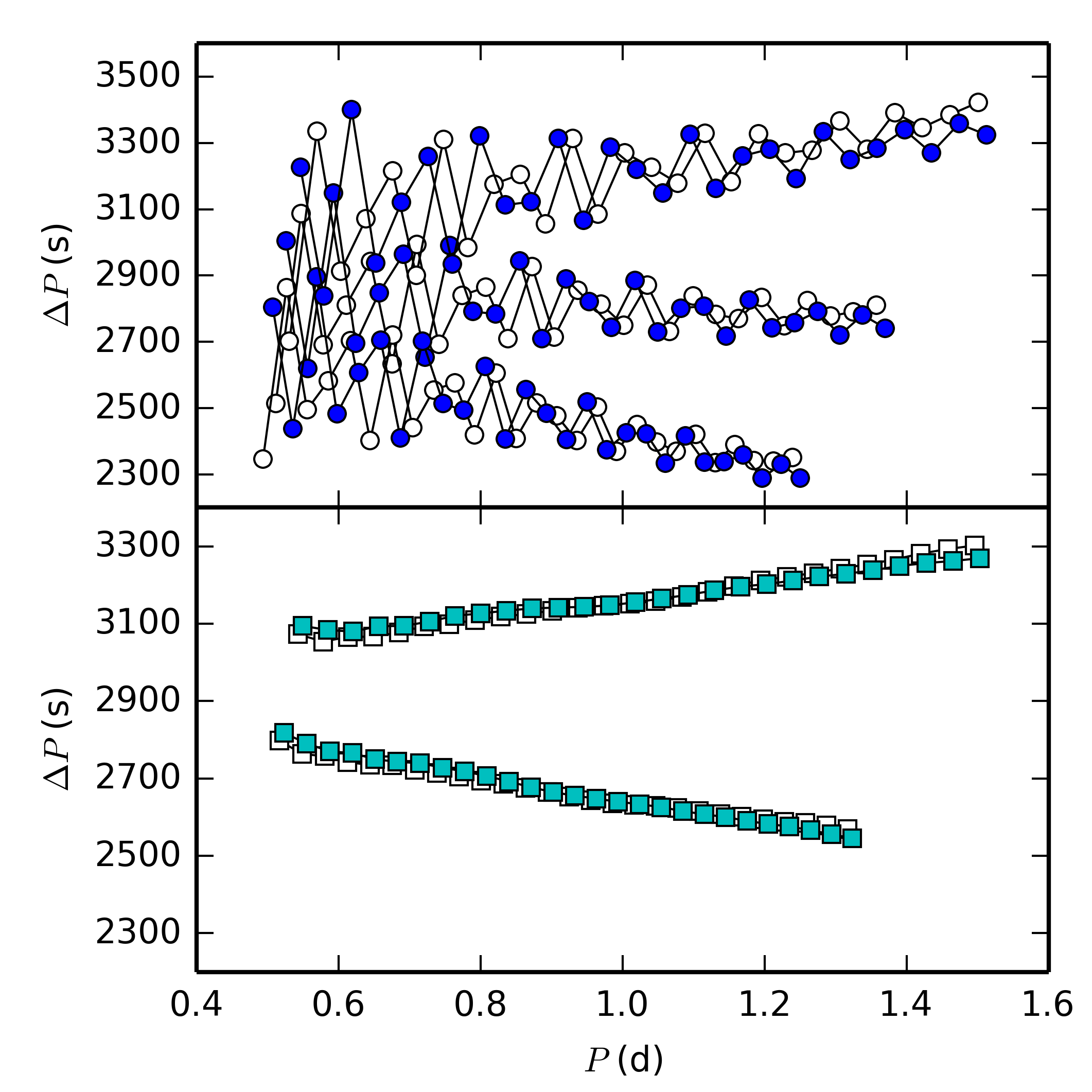}}
\subfloat[Period spacing for Model 3 and Model 5.]{\label{fig:gmodes_ledoux_tar:b}\includegraphics{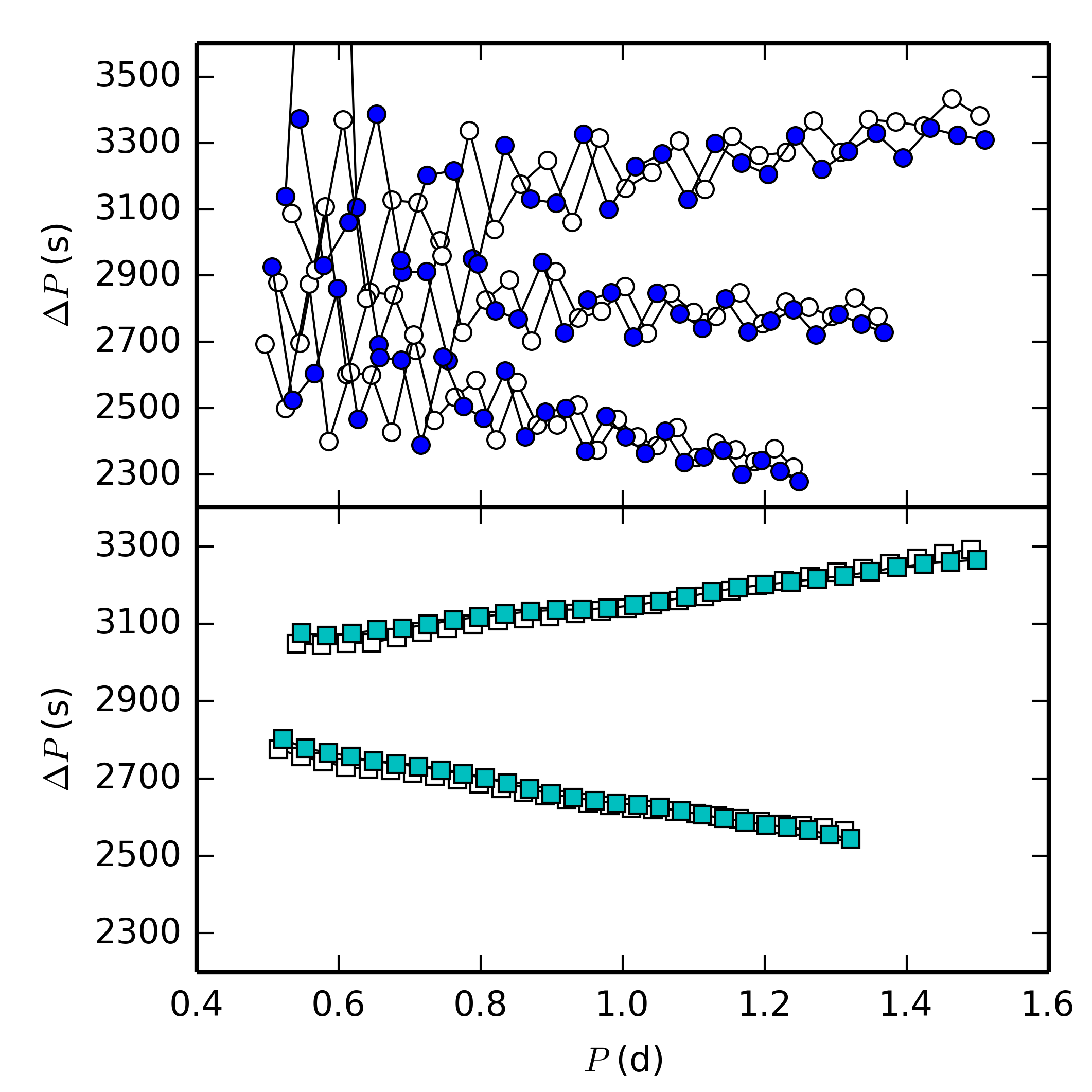}}
\caption{Difference between first-order perturbation approximation and TAR for the rotation rates $\Omega_{c,1}=0.13963~\mathrm{d}^{-1}$ and $\Omega_{c,2}=0.09034~\mathrm{d}^{-1}$. GYRE zonal modes with a Ledoux shift applied are shown as the empty symbols and GYRE modes calculated using the TAR are shown as the filled symbols, where the primary is in dark blue and the secondary in light blue.}
\label{fig:gmodes_ledoux_tar}
\end{figure*}

\begin{figure*}
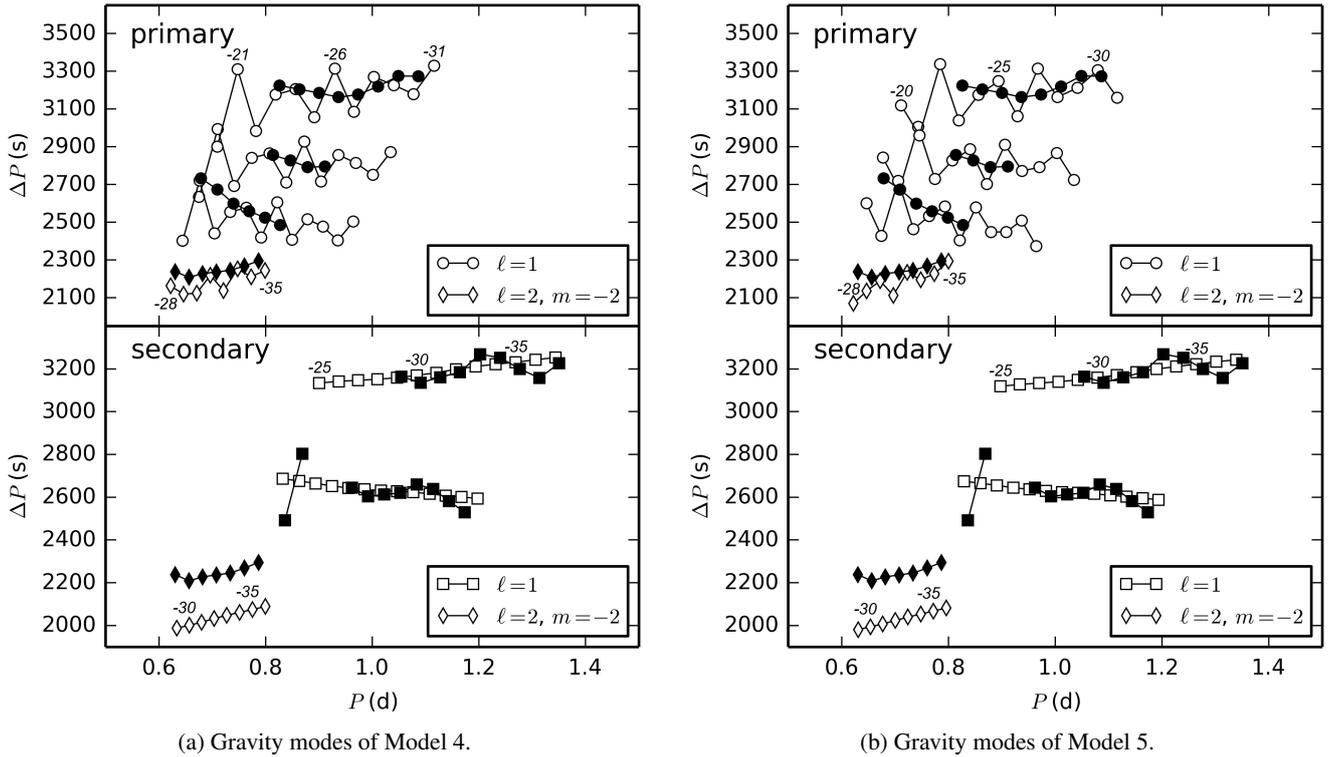

\centering
\subfloat[Gravity modes of Model 4.]{\label{fig:gmodes_model_tar:a}\includegraphics{MESAexpov_gyre_periodspacing_lowchi2.png}}
\subfloat[Gravity modes of Model 5.]{\includegraphics{MESAstepov_gyre_periodspacing_lowchi2.png}}
\caption{Same as Fig.~\ref{fig:gmodes_model_lowchi2}, but using the TAR instead of the Ledoux splitting to estimate the influence of rotation.}
\label{fig:gmodes_model_tar}
\end{figure*}

To test the accuracy of the rotation rates derived with the Ledoux splitting, i.e.\ $\Omega_{c,1}=0.13963$ and $\Omega_{c,2}=0.09034$, we recompute all modes using the TAR. At the rotation rates that were found from the first-order perturbations, we find a disagreement between the GYRE modes and the observations. However, we find that by altering the rotation rates, we cannot obtain a better fit. Rather, we find that the observed shift is due to a difference in the mean period spacing values. We selected Models 2 and 3 as having the lowest $\chi^2$ for the asymptotic period spacing, which is only valid for zonal modes in a non-rotating case. As the Coriolis force shifts the zonal modes to higher frequencies, we find a decreased period spacing. Therefore, we find a better fit, when selecting younger models (Models 4 and 5; see Table~\ref{tab:bestmodels}) where $\Delta P_{c,1}$ is about 64~s higher and $\Delta P_{c,2}$ is about 36~s higher, for both the exponential and step overshooting grids. Figure~\ref{fig:gmodes_ledoux_tar} shows how the period spacings of the best models from the grid search and these younger models compare, using the two different approximations (Ledoux versus TAR) for the puslations. We compare the period spacings of these younger models to the observations in Fig.~\ref{fig:gmodes_model_tar}. As can be seen, there is only a slight difference between Fig.~\ref{fig:gmodes_model_tar} and Fig.~\ref{fig:gmodes_model_lowchi2}, which mainly results from adopting younger models.

\section{Discussion}
\label{sec:discussion}

\subsection{On the mixing parameters of the coeval models}

\begin{figure}[t!]
\centering
\includegraphics{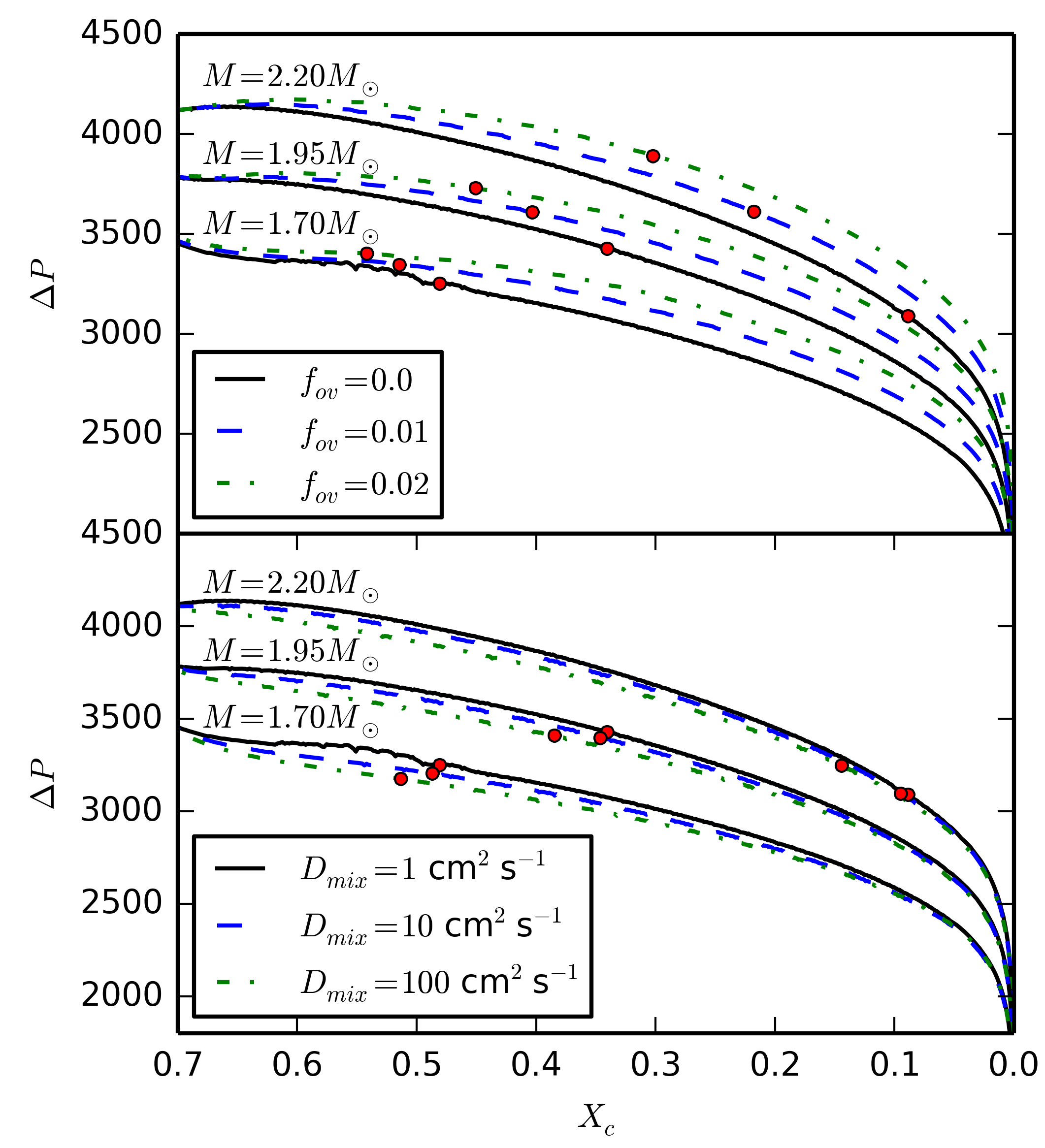}
\caption{Asymptotic period spacing $\Delta P$ as a function of central hydrogen mass fraction $X_c$ for different masses and with different values of $f_\mathrm{ov}$ (top panel) and $D_\mathrm{mix}$ (bottom panel). The red points denote ages of 0.6 Gyr.}
\label{fig:ps_ov_dmix}
\end{figure}

During the analysis of all three grids of coeval models based on $\Delta P_c$, $T_\mathrm{eff}$, and $R$, we found a clear trend for high overshooting and low diffusive mixing for the primary, and low overshooting and high diffusive mixing for the secondary. The main reasons why these parameters influence our model selection is that stronger mixing increases the main sequence lifetime of a star by providing fresh hydrogen to the nuclear-burning core. By directly affecting the stellar structure they also alter the Brunt-V\"ais\"al\"a frequency and thus $\Delta P_c$. Both effects are illustrated in Fig.~\ref{fig:ps_ov_dmix}. In this figure it can also be seen that higher mass models have a higher period spacing at a given $X_c$. Yet, the evolution of $\Delta P_c$ is faster for the primary and drops to the observed value in shorter time than for the secondary, when the strength of mixing is equal in both components. Thus, a higher value of $f_\mathrm{ov}$ for the primary is needed to slow the evolution of $\Delta P_c$ down. Additionally, it extends the evolutionary track to reach the observed $T_\mathrm{eff,1}$, when it would otherwise turn off before the observed error box. For the secondary, additional mixing is mainly required to adjust the evolution of $\Delta P_c$ with that of the primary. As models with $f_\mathrm{ov}=0.0$ and $\log(D_\mathrm{mix})=0.0$ already cross the observational error box on the HRD, increasing $f_\mathrm{ov}$ would result in a too cool and too big star. Diffusive mixing, on the other hand, has a weaker influence on the main sequence lifetime than overshooting and is thus the preferred mixing mechanism for the secondary. In summary we state that the equal age requirement and the stellar parameters from binary modelling, serve as a stringent constraint to derive the width of the core-overshooting region and the strength of radial envelope mixing in both stars at the current evolutionary stage. Variations throughout the stellar evolution of these parameters are likely to occur. Both stars could thus have had similar values near the ZAMS or started off with different values at birth.

\subsection{The period spacing morphology}

\begin{figure*}
\centering
\subfloat[$M=1.81~M_\odot$, $Z=0.0125$, $f_\mathrm{ov}=0.011$, $\log(D_\mathrm{mix})=0.75$]{\includegraphics{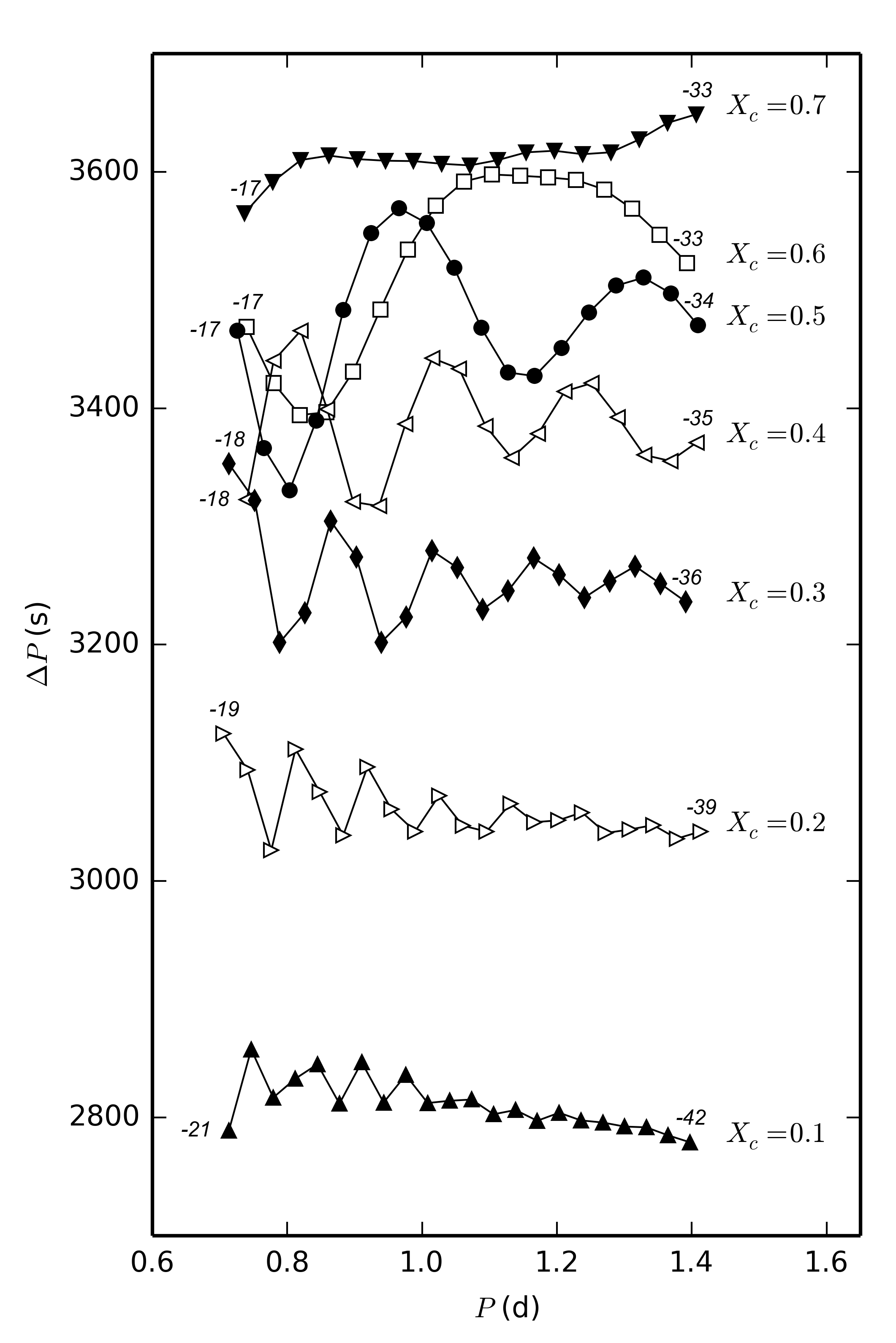}}
\subfloat[$M=1.81~M_\odot$, $Z=0.0125$, $\alpha_\mathrm{ov}=0.11$, $\log(D_\mathrm{mix})=0.75$]{\includegraphics{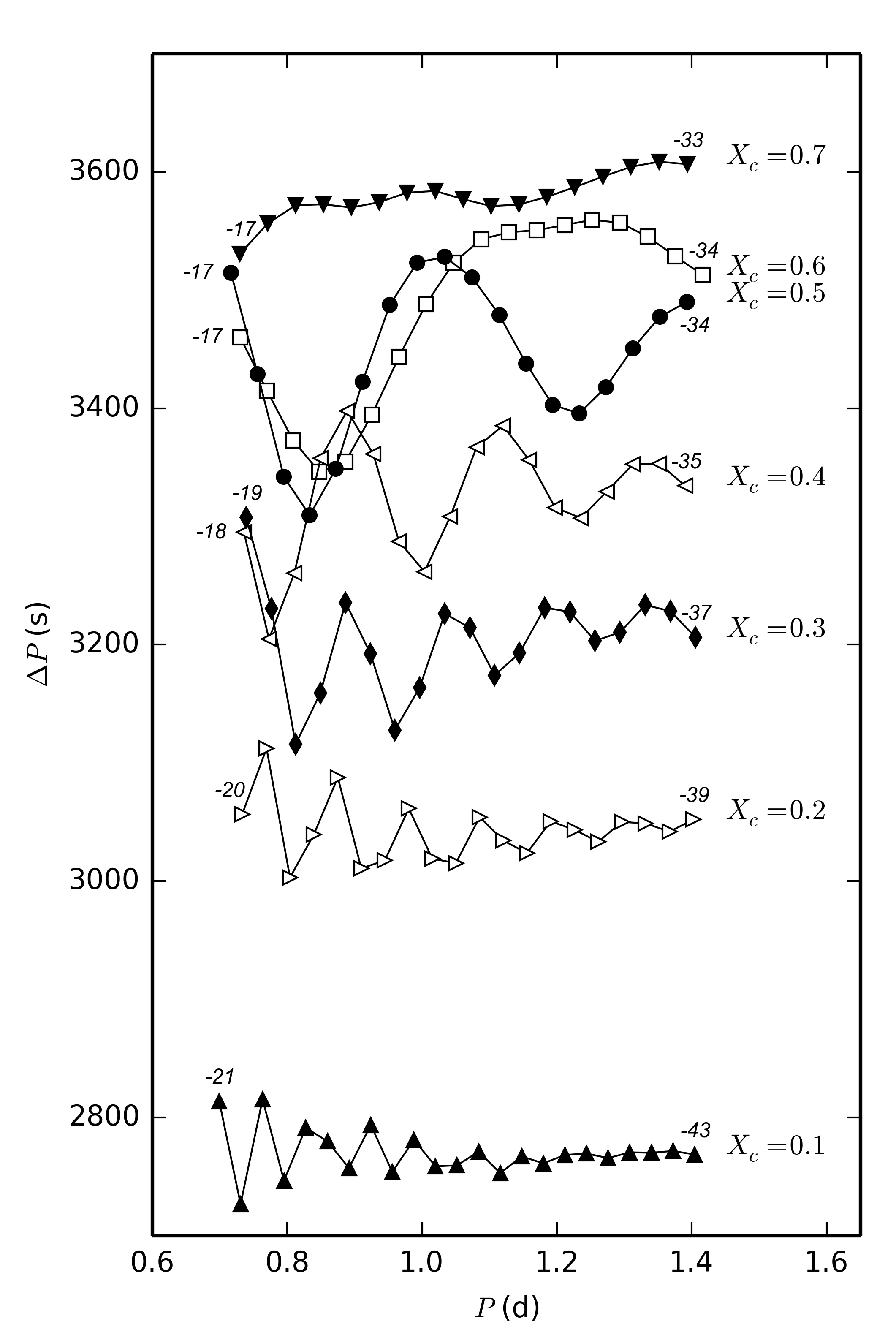}}
\caption{Period spacing for different values of $X_c$ along two evolutionary tracks incorporating exponential core-overshooting (left panel) or step overshooting (right panel). The radial orders of the first and last mode in the period spacing is given. The filling of the symbols is alternated to enhance visibility.}
\label{fig:ps_track}
\end{figure*}

\begin{figure*}
\centering
\subfloat[$M=1.76~M_\odot$, $Z=0.0125$, $f_\mathrm{ov}=0.005$, $X_c=0.20$]{\includegraphics{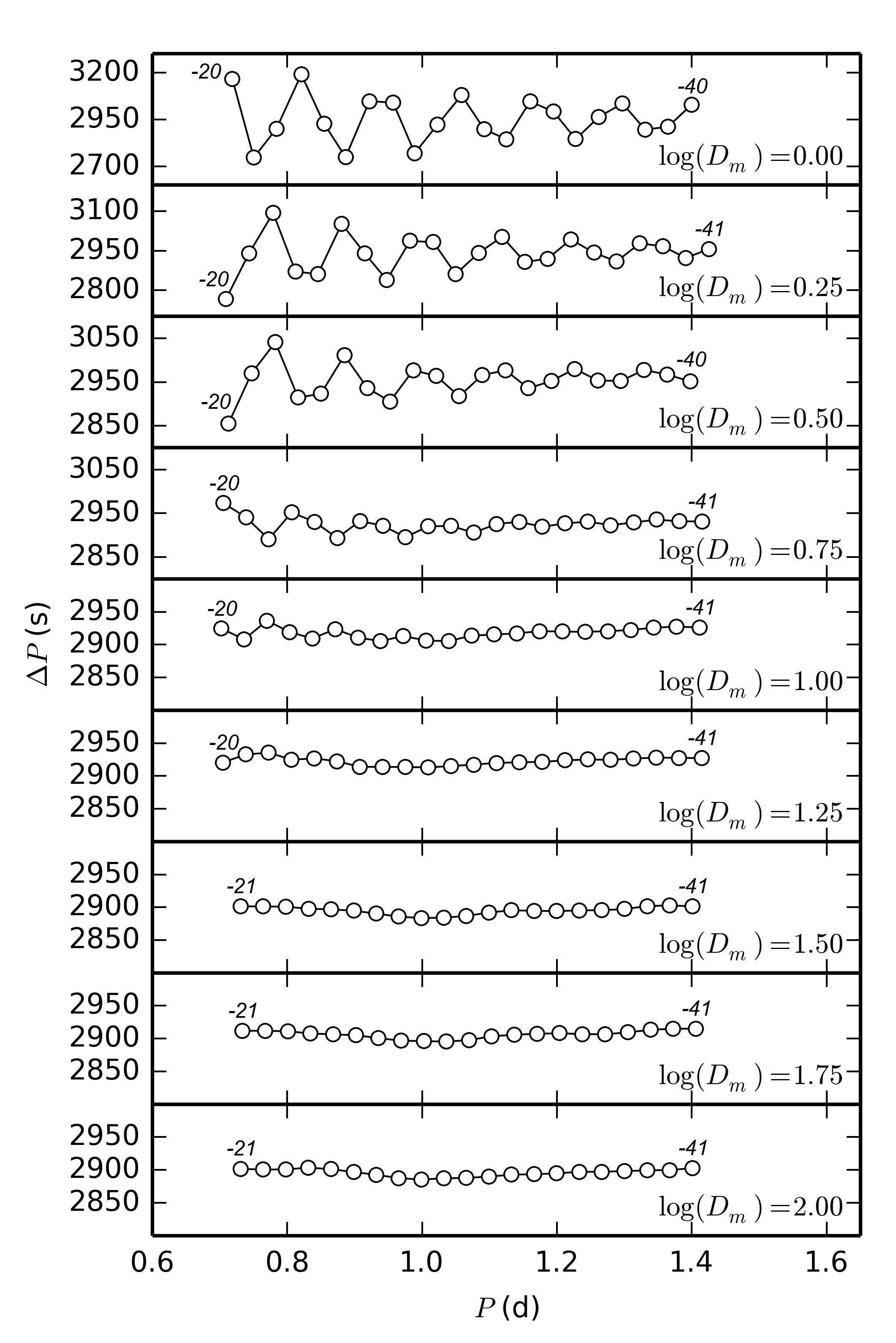}}
\subfloat[$M=1.76~M_\odot$, $Z=0.0125$, $\alpha_\mathrm{ov}=0.05$, $X_c=0.21$]{\includegraphics{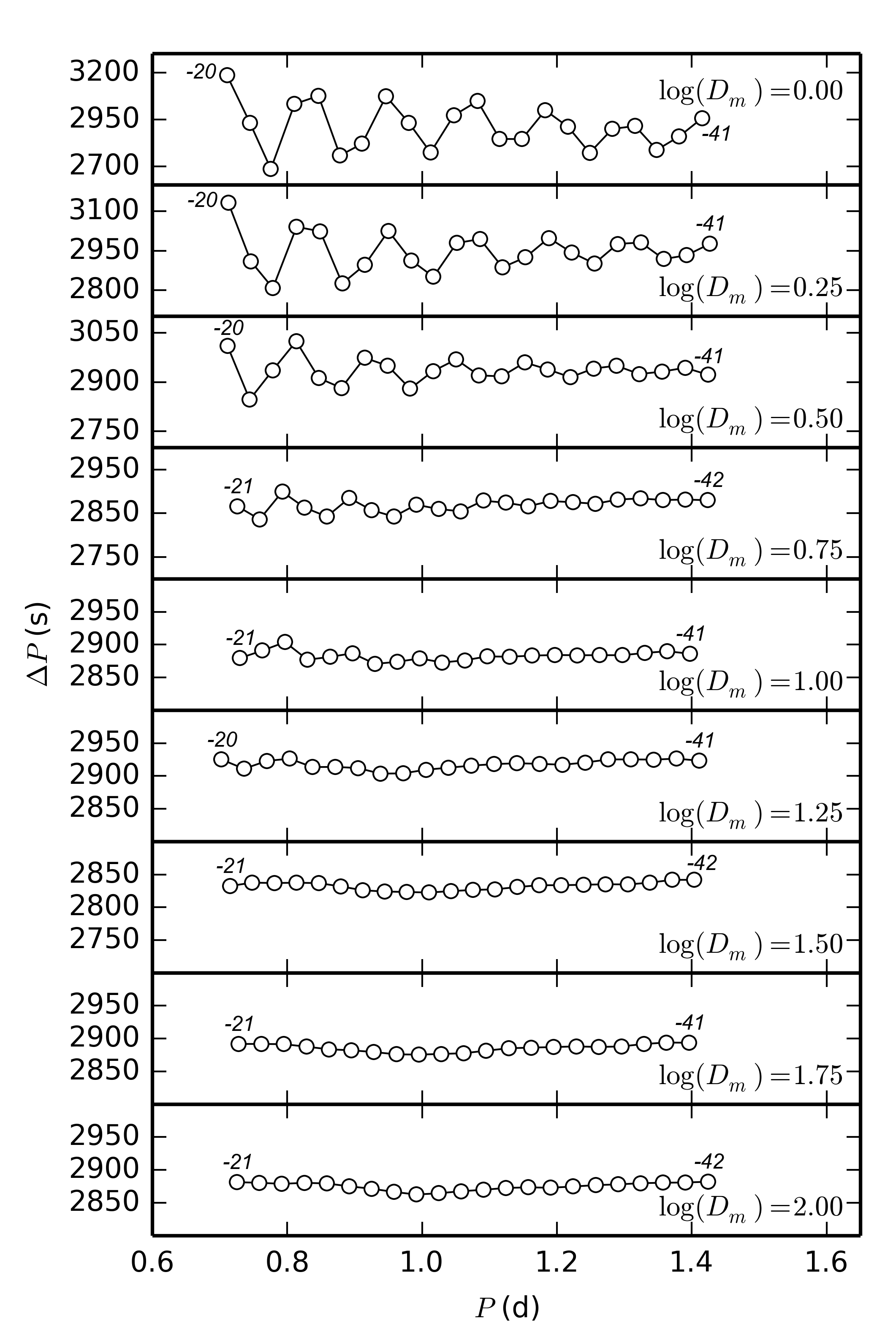}}
\caption{Period spacing for different values of $\log(D_\mathrm{mix})$ along two evolutionary tracks, one with exponential core-overshooting (left panel) and one with step overshooting (right panel). The radial orders of the first and last mode in the period spacing is given. }
\label{fig:ps_dmix}
\end{figure*}

When focusing only on the morphology of the period spacing we find Models 2 and 3 to be less satisfying fits to the observations than suggested by the mean period spacing values. While the computed period spacing for the primary has variations that are too strong and too short, the structure of the secondary model is too smooth (see Figs.~\ref{fig:gmodes_model_lowchi2} and \ref{fig:gmodes_model_tar}). \citet{Miglio2008} showed that the interval of the dips in period spacing depends on the steepness of the chemical composition gradient near the core, which grows with increasing age or decreasing $X_c$ as more and more hydrogen is burned into helium. In Fig.~\ref{fig:ps_track} we show how the period spacing evolves for different values of $X_c$ for a model of $M=1.81~M_\odot$. For the secondary, the lack of structure in the computed period spacing can be explained by high diffusion, which diminishes $\nabla_\mu$ near the core. In Fig.~\ref{fig:ps_dmix} it is clearly visible how the dips in period spacing disappear with increasing $\log(D_\mathrm{mix})$. Hence, the period spacing morphology points to stars with a higher $X_c$ than we found with the grid search, explained in Sect.~\ref{sec:modelling}. Indeed, we find a better fit with models where $X_c\approx0.35$ and $0.5\leq\log(D_\mathrm{mix})\leq0.75$, while we find that overshooting only has weak influence. The similarity of these values for both stars is a direct consequence of the similar morphology in the observed period spacings of both components, however, they are in contradiction to the other observational constraints. By increasing $X_c$ the stars move closer to the ZAMS and outside the error box defined by $T_\mathrm{eff}$ and $R$ from binary modelling. In even stronger disagreement, however, are the asymptotic period spacings, as they are several $\sigma$ above the observed values.

\begin{figure*}
\centering
\subfloat[Gravity modes of Model 6.]{\label{fig:gyre_model_structure:a}\includegraphics{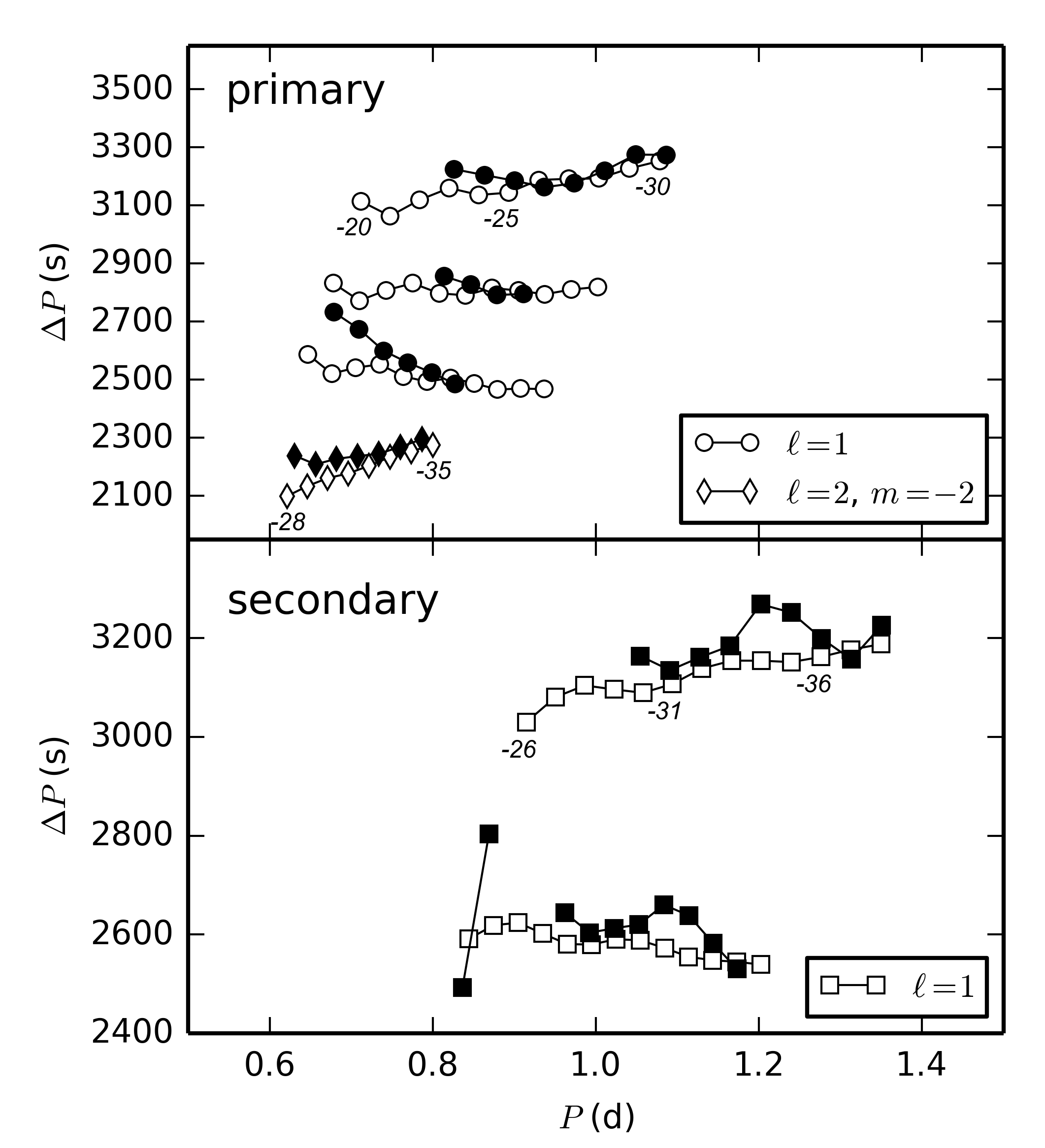}}
\subfloat[Pressure modes of Model 6.]{\label{fig:gyre_model_structure:b}\includegraphics{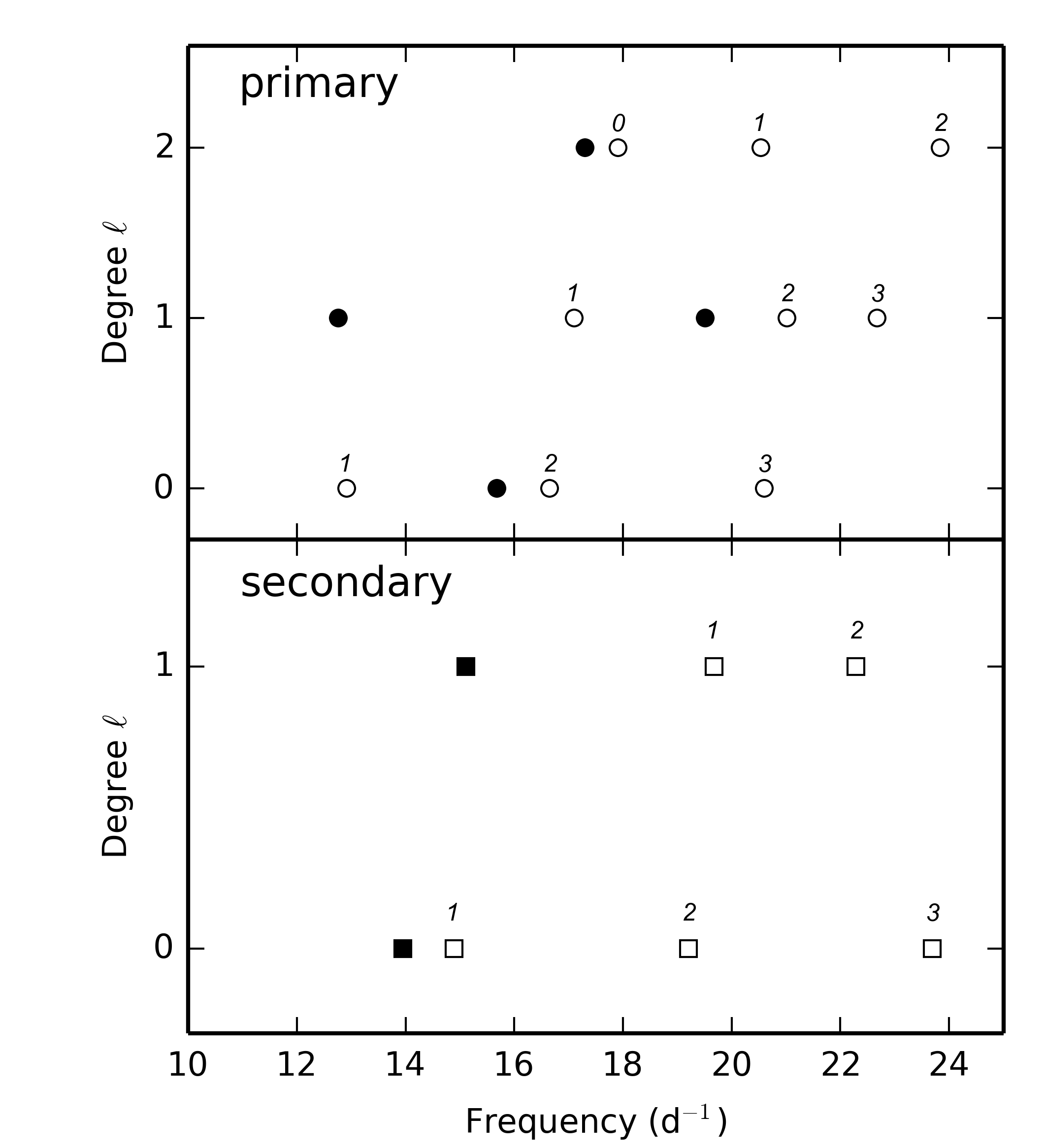}}
\caption{Comparison of theoretical modes (empty symbols) to observed modes (filled symbols) for the primary (top panels) and the secondary (bottom panels). The radial orders are given near selected modes.}
\label{fig:gyre_model_structure}
\end{figure*}

\begin{figure}
\centering
\includegraphics{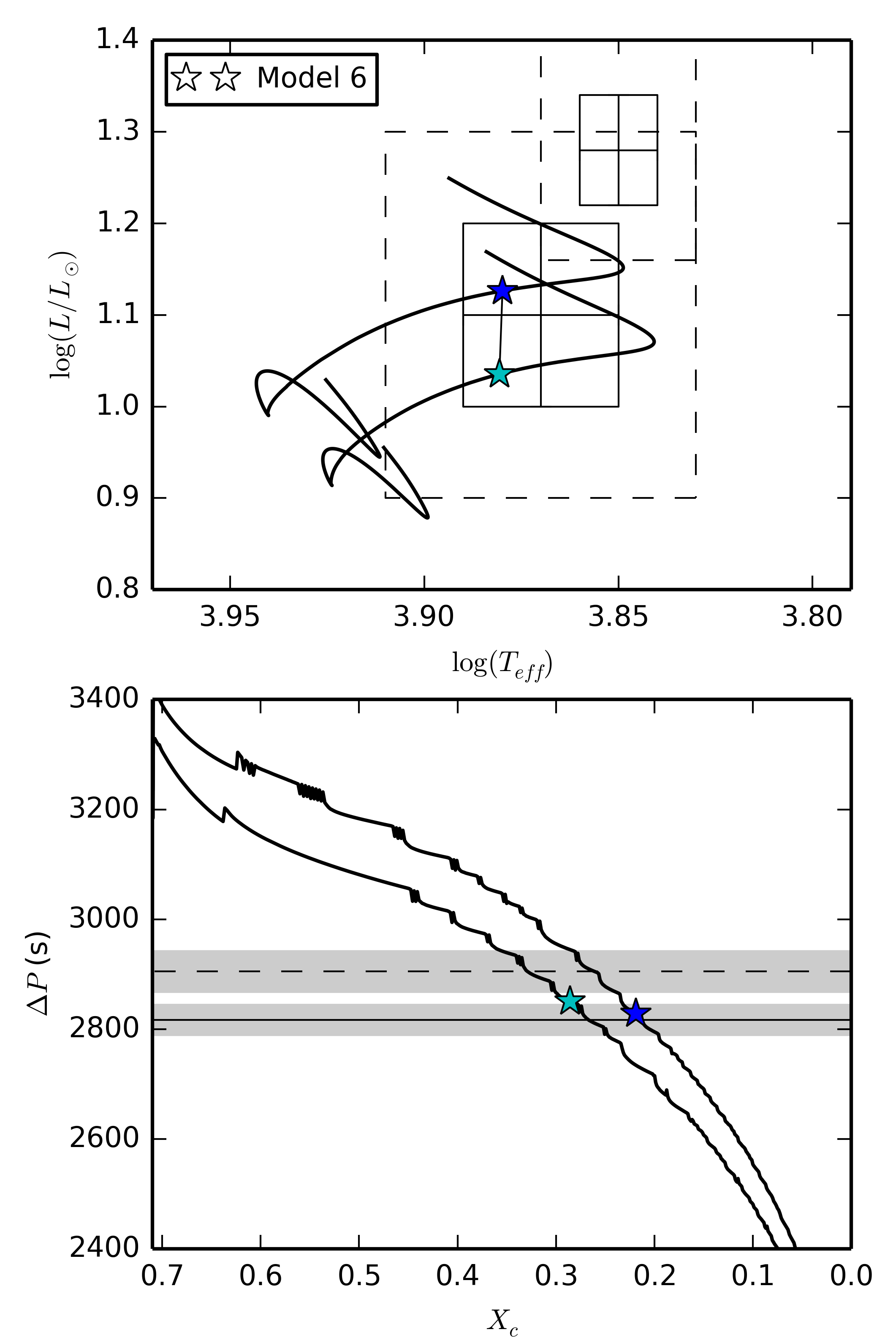}
\caption{Same as Fig.~\ref{fig:HRDbest} but for Model 6, indicated by stars and its evolutionary track as the black solid line. Here the $2\sigma$ areas for $T_\mathrm{eff,1,2}$ and $\log(L_{1,2}/L_\odot)$ are also shown as dashed boxes.}
\label{fig:HRDstructure}
\end{figure}

To obtain better agreement between the period spacing morphology and the mean period spacing values, as well as the stellar parameters, we exploit correlations between mass, metallicity, and $\Delta P_c$. For a given $X_c$, the asymptotic period spacing decreases with decreasing mass and decreasing metallicity. Moreover, models with low metal content are more luminous than those with high metal content and, therefore, mimic higher mass models with high metallicity in the HRD. To decrease $\Delta P_c$ as much as possible, we use a secondary mass of $M_2-3\sigma_{M_2}=1.6~M_\odot$ and determine the primary mass via the mass ratio as $M_1=M_2/q=1.67~M_\odot$. We choose a metallicity of $Z=0.01$, which is still in agreement with the slightly sub-solar metallicity reported by \citet{Schmid2015}. The mixing parameters were chosen to conform with the observed structure in the measured period spacing pattern. From the computed evolutionary tracks we determine the coeval Model 6, where $\chi^2_\mathrm{tot}$ is at a minimum (all grid and global parameters of Model 6 can be found in Table~\ref{tab:bestmodels}). For this model, we compute the zonal dipole and quadrupole modes with GYRE and apply the Ledoux splitting for the prograde and retrograde modes. The derived rotation rates for this model, are $\Omega_{c,1}=0.13969\pm0.00003~\mathrm{d}^{-1}$ and $\Omega_{c,2}=0.09047\pm0.00004~\mathrm{d}^{-1}$. We show the result for this test in Fig.~\ref{fig:gyre_model_structure}, where the computed g and p~modes are compared to the observed modes. We find a much better fit of the period spacing morphology and even the asymptotic period spacing values lie within $2\sigma$ of the observed mean values (Fig.~\ref{fig:gyre_model_structure:a}). Some room for improvement exists for the global parameters of the primary model, as can be seen in Fig.~\ref{fig:HRDstructure}. The $T_\mathrm{eff}$ and $\log(L/L_\odot)$ of the model still lie within $3\sigma$ of the observed values, but its radius $R=2.12~R_\odot$ lies outside this range.

\subsection{Varying the He mass fraction}

In the model grids discussed above, we have assumed an initial hydrogen mass fraction $X_i=0.71$. Varying $X_i$ and thereby adapting the initial helium mass fraction $Y_i$ influences the positions of the tracks on the HRD and the asymptotic period spacing. Increasing $Y_i$, while decreasing $X_i$ causes a star of the same mass and $Z$ to become more luminous. Furthermore, for a given $X_c$, $\Delta P_c$ increases. We therefore tested whether changing the helium mass fraction would lead to higher masses that are in better agreement with the values from binary modelling, following the same modelling procedure as carried out for fixed $X_i=0.71$. We computed two additional small grids of models for $X_i=0.7$ and $X_i=0.72$, using exponential overshooting. The best fitting model with $X_i=0.72$ has a primary mass of $M_1=1.83~M_\odot$, which is slightly higher than the mass of Model~2. However, this increase in mass is accompanied with a lower metallicity of $Z=0.0115$. The secondary mass is the same as for Model~2, while the overshooting and diffusion parameters only change slightly ($f_\mathrm{ov,1}=0.01$, $\log D_\mathrm{mix,1}=0$, $f_\mathrm{ov,2}=1.75$, and $\log D_\mathrm{mix,2}=1.75$).

\subsection{The p modes}

The observed p~modes provide weaker constraints for both components than the g~modes because of their uncertain mode identification, which is based on the assumption that all observed multiplets are complete. Models 2 and 3, which best fit the observed mean period spacing and binary parameters, confirm the mode identification for the secondary (Fig.~\ref{fig:pmodes_model}). The singlet and highest amplitude p~mode $f_{s,2}$ fits the first overtone mode better than the radial fundamental. For the primary, the situation is somewhat more ambiguous. For the singlet $f_{s,1}$ and the triplet centre frequency $f=19.51028~\mathrm{d}^{-1}$, neither $\ell$ nor $k$ can be assigned without doubt. As for the g~modes, it is not possible to distinguish between exponential and step overshooting using the p~modes.

\citet{Schmid2015} used the rotational splittings observed in the p~modes and the g~modes to provide an estimate of the core-to-surface rotation rates. They showed that both stars might have a slight differential rotation in the radial direction. We now use our best models to compute the Ledoux constant $C_{n,\ell}$ and derive the envelope rotation rates based on the p-mode splittings. Since the splittings detected by \citet{Schmid2015} show a significant asymmetry, which cannot be reproduced by the first-order Ledoux approximation, we use the average $\Delta f$ per multiplet.

For the triplet associated with the secondary (see Fig.~\ref{fig:pmodes_observed}) we find that $\beta_{n,\ell}=(1-C_{n,\ell})$ is close to 1, which is the expected value for pure p modes. This yields an envelope rotation rate of $\Omega_{e,2}=0.122\pm0.002~\mathrm{d}^{-1}$ for the secondary. Thus, we observe differential rotation in the radial direction with a slower core than envelope, given the ratio of surface-to-core rotation rate $\Omega_{e,2}/\Omega_{c,2}=1.35\pm0.02$. This value is equal for Models 2 and 3.

For the primary, the situation is slightly more complicated. Using $\beta_{n,\ell}$ for the modes of radial orders $n=1$ and $n=4$ for the dipole modes and $n=1$ for the quadrupole mode, we derive rotation rates in the range $0.137\leq \Omega_{e,1}~(\mathrm{d}^{-1}) \leq 0.173$. The envelope of the primary can thus be either faster or slower than its core, depending on the p-mode multiplet. When we examine the $\beta_{n,\ell}$ values of the individual modes more closely, we find that some have values below 0.9. For the mode near $f=19.51~\mathrm{d}^{-1}$ we find $\beta_{4,1}\approx0.8$ and a rotation rate of $0.16-0.17~\mathrm{d}^{-1}$, while $\beta_{3,1}\approx0.97$ and hence $\Omega_{e,1}\approx0.135~\mathrm{d}^{-1}$. This is because the $n=4$ mode has a mixed mode character and not only probes the envelope (like pure p~modes) but also has a high amplitude near the convective core (like pure g~modes). This is also the case for the quadrupole mode, where we find envelope rotation rates of $0.14-0.15~\mathrm{d}^{-1}$. To obtain a better estimate of the surface-to-core rotation rate, we only use the rotation rates derived from the pure p-mode triplet near $f=12.76~\mathrm{d}^{-1}$. For this triplet $\Omega_{e,1}$ ranges from $0.131\pm0.003~\mathrm{d}^{-1}$ to $0.137\pm0.003~\mathrm{d}^{-1}$, using either modes of $n=1$ or $n=2$ of both Models~2 and 3. Thus, we find almost uniform rotation with a slightly faster core than envelope, with $\Omega_{e,1}/\Omega_{c,1}=0.94\pm0.2$ up to $0.98\pm0.2$.

Moreover, we use the p~modes to compare the scenario that predicts the mean period spacing and the binary parameters (Models 2 and 3) to the scenario that predicts the period spacing morphology (Model 6). We find that the p~modes of Model 6 have frequencies that are too high and therefore do not agree with the observations. Aside from the binary parameters, the p-mode frequencies are thus another indication that Models 2 and 3 are indeed better representations for the two stars of KIC\,10080943 than Model 6.

\section{Summary}

In this paper we have presented detailed seismic modelling for two F-type g- and p-mode hybrid pulsators that reside in the binary system KIC\,10089043 \citep{Keen2015,Schmid2015}. We calculated stellar model grids with MESA, covering the observed range of stellar parameters, and compared the observed pulsation modes to predictions by the stellar pulsation code GYRE. For the analysis, we were able to exploit observational constraints from the binarity and the pulsations. This was implemented by requiring an equal age and composition, the mass ratio and the position in the HRD within $2\sigma$ of the binary solution, together with asymptotic period spacing values within $2\sigma$ of the observed mean period spacing. With this approach, we provide the first such consistent seismic modelling of a binary F-type g-mode pulsator. We find that the stars have an age of $\simeq1.1$~Gyr and have a low central hydrogen mass fraction of $X_{c,1}\simeq0.1$ and $X_{c,2}\simeq0.2$. This is a consequence of the observed mean period spacing values, which are below 3000~s and, thus, rather low for $2~M_\odot$ stars. In fact, our best models have masses $2\sigma$ below the binary values. However, the age is in good agreement with the observed HRD positions of both stars with cool temperatures and big radii. Furthermore, we were able to determine the rotation rates of the regions near the core to be $\Omega_{c,1}=0.13963\pm0.00007~\mathrm{d}^{-1}$ and $\Omega_{c,2}=0.09034\pm0.00002~\mathrm{d}^{-1}$. When the traditional approximation of rotation is used at these rotation rates instead of the first-order perturbative Ledoux approximation, a lower asymptotic period spacing is required for both stars. This is because the zonal modes are influenced by the Coriolis force and shifted to higher frequencies.

To test the shape of the convective-core overshooting we computed two grids that differ only in the overshooting prescription. One grid uses exponential core overshooting, while the other uses step overshooting. We could not find a significant difference between these two grids, in the sense that the difference in morphology of the period spacing pattern between these two descriptions is by far inferior to the discrepancy between the observed and theoretical morphology. We could show, however, that the stellar parameters derived from binary modelling and the requirement of equal age and composition for both stars, can constrain the amount of overshooting and diffusive mixing. For the primary, we find high values of overshooting and low values of diffusion and for the secondary we find the opposite. The best model for the exponential overshooting grid is $M_1=1.82~M_\odot$, $M_2=1.76~M_\odot$, $Z=0.0125$, $f_\mathrm{ov,1}=0.008$, $f_\mathrm{ov,2}=0.005$, $\log(D_\mathrm{mix,1})=0.25$, and $\log(D_\mathrm{mix,2})=1.75$; $f_\mathrm{ov}$ is expressed in local pressure scale height $H_P$ and $D_\mathrm{mix}$ has a unit of cm$^2$\,s$^{-1}$. For the step overshooting grid, the best model is similar at $M_1=1.81~M_\odot$, $M_2=1.76~M_\odot$, $Z=0.0125$, $\alpha_\mathrm{ov,1}=0.11$, $\alpha_\mathrm{ov,2}=0.05$, $\log(D_\mathrm{mix,1})=1.25$, and $\log(D_\mathrm{mix,2})=1.5$.

These best models fail to predict the detailed period spacing structure, which points to the limitations of 1D stellar models in terms of chemical mixing in the radial direction only. For these 1D stellar models, we find a better agreement with the observed structures for stars that have a higher $X_c$ and more similar mixing parameters, which contradicts the observed mean period spacing and stellar parameters. We are able to find a coeval model with low masses and metallicity that fits both the morphology and mean value of the period spacing. Yet, the p-mode frequencies of this model are too high to reproduce the observed frequencies. This points to mean densities that are too high or radii that are too small. This model, indeed, is discrepant with the binary modelling at $+3\sigma$ level. Especially, the primary radius of the model is $4\sigma$ below the observed value. An independent measurement of the radius, for example through a precise distance provided by the ESA Gaia mission \citep{Perryman2001}, could help to firmly exclude such younger models. The fact that the variations of the period spacing of the primary are too strong clearly points towards missing extra mixing in the transversal direction, which is perhaps of rotational origin.

During the analysis for this paper, we ignored the influence both stars have on each other and modelled them as coeval single stars. This approach is  justified, as the observed interaction is weak and we relied on the g~modes, whose probing power is situated near the core. We can estimate the tidal influence by the tide-generating potential, which depends on the small parameter $\varepsilon_T=(R_1/a)^3(M_2/M_1)$ \citep[e.g.][]{Willems2002} and find $\varepsilon_T\approx3.7\times10^{-4}$ for KIC\,10080943. The influence of the centrifugal force due to the rotation on the pulsations can be estimated by $(\Omega/f_\mathrm{co})^2$, which lies between 0.01 and 0.017 for the primary and between 0.005 and 0.014 for the secondary \citep[see e.g.][]{Aerts2002}. Thus, the centrifugal force has a much stronger influence on the pulsations than the tides, but is so small that our approach based on the Coriolis force above is fully justified. Also, the brightening signal, which is detected in the high-precision \textit{Kepler} light curve, has a small amplitude and the reflection effect outweighs the contribution of the tidal distortion. Some evidence of the tidal forces, however, could be detected in one $\ell=1$ g mode, which has a frequency of exactly $15f_\mathrm{orb}$, and in the rotational splitting of the p~modes \citep{Schmid2015}. The next step towards improving the modelling of this binary would be to create a binary model that takes these interactions into account. Such models could also help gauge the effect the tidal forces have on the level of mixing or the convective-core size. The best models we obtained so far constitute a fruitful starting point for more detailed studies such as these. Other future work could include mode excitation. As of now, GYRE does not include a pulsation-convection interaction, which would be necessary to study the mode excitation by convective flux blocking \citep{Guzik2000,Dupret2005}. Therefore, KIC\,10080943 is a rich target, which still holds potential for improving stellar and seismic models.

\begin{acknowledgements}
The research leading to these results received funding from the European Research Council (ERC) under the European Union's Horizon 2020 research and innovation programme (grant agreement N~670519: MAMSIE). The computational resources and services used in this work were provided by the VSC (Flemish Supercomputer Center), funded by the Hercules Foundation and the Flemish Government -- department EWI. We would like to thank Ehsan Moravveji, P\'eter I. P\'apics, and Timothy Van Reeth for fruitful discussions and valuable input, and the anonymous referee for comments that helped improve this manuscript.
\end{acknowledgements}

\bibliographystyle{aa} 
\bibliography{bibliography}

\begin{appendix}

\section{MESA input file}
\label{app:inlist}

\begin{verbatim}
! inlist for KIC10080943 grid calculations
! evolve main sequence model starting from ZAMS
! Parameters input via run_star_extras.f 
! not in inlist
!
! Input via run_star_extras.f: 
! new_Z, new_Y, initial_mass, min_D_mix, and
! step_overshoot_f_above_burn_h or
! overshoot_f_above_burn_h

&star_job

  ! history and profile columns
    history_columns_file = 'hist.list'
    profile_columns_file = 'prof.list'

!!! begin with a pre-main sequence model
    create_pre_main_sequence_model = .false.
    
!!! start from ZAMS:
    relax_Z = .true.
    change_Z = .true.
    relax_initial_Z = .true.
    change_initial_Z = .true.
    ! new_Z set via run_star_extras.f
    
    relax_Y = .true.
    change_Y = .true.
    relax_initial_Y = .true.
    change_initial_Y = .true.
    ! new_Y set via run_star_extras.f

!!! Composition and opacities (Asplund09+OPAL):    
    kappa_file_prefix = 'a09'
    kappa_lowT_prefix = 'lowT_fa05_a09p'
    initial_zfracs = 6
    
    change_lnPgas_flag = .true.
    change_initial_lnPgas_flag = .true.
    new_lnPgas_flag = .true.
    
    change_net = .true.
    new_net_name = 'pp_cno_extras_o18_ne22.net'
    change_initial_net = .true.

  ! display on-screen plots
    pgstar_flag = .false.

/ !end of star_job namelist


&controls

!!! Starting specifications
    ! initial_mass set via run_star_extras.f

!!! Atmosphere
    which_atm_option = 'photosphere_tables'


!!! Mixing
  ! Diffusive mixing
  ! Minimal mixing in radiative zone
    set_min_D_mix = .true.
    ! min_D_mix set via run_star_extras.f
    remove_small_D_limit = 0
    
  ! Convection MLT
    mixing_length_alpha = 1.8
        
    use_Ledoux_criterion = .false.
    
  ! Over- and undershooting
    D_mix_ov_limit = 5d-2
    
    overshoot_f0_above_burn_h = 0.001
    
  ! Exponential overshooting and coarse grid
    ! overshoot_f_above_burn_h set 
    ! via run_star_extras.f
    
  ! Step function overshooting grid 
    ! step_overshoot_f_above_burn_h set 
    ! via run_star_extras.f
    
    step_overshoot_D = 0
    step_overshoot_D0_coeff = 1
    
    num_cells_for_smooth_brunt_B = 0 
    num_cells_for_smooth_gradL_composition_term = 0


!!! Time step control
    varcontrol_target = 2.5d-5
    
    max_age = 2d10
    max_years_for_timestep = 1d7
  
  ! Stop condition
  ! stop when the center mass fraction of h1 
  ! drops below this limit
    xa_central_lower_limit_species(1) = 'h1'
    xa_central_lower_limit(1) = 1d-3
    when_to_stop_rtol = 1d-3
    when_to_stop_atol = 1d-3


!!! Mesh grid
    max_allowed_nz = 10000
    mesh_delta_coeff = 0.5
    cubic_interpolation_in_Z = .true.
    
    xa_function_species(1) = 'he4'
    xa_function_weight(1) = 100
    xa_function_param(1) = 1d-2
         
    xa_function_species(2) = 'he3'
    xa_function_weight(2) = 100
    xa_function_param(2) = 1d-5
    
    mesh_dlogX_dlogP_extra(:) = 0.1         ! resol coeff for chemical gradients
    mesh_dlogX_dlogP_full_on(:) = 2
    mesh_dlogX_dlogP_full_off(:) = 1
      
    mesh_logX_species(1) = 'he4'            ! taking into account abundance of He4
    mesh_logX_min_for_extra(1) = -6         ! for abundances larger than ~
    mesh_logX_species(2) = 'n14'            ! taking into account abundance of N14
    mesh_logX_min_for_extra(2) = -6
    mesh_logX_species(3) = 'c12'
    mesh_logX_min_for_extra(3) = -6
    
    xtra_coef_czb_full_on = 1.0             ! use this coef if center_he4 below this
    xtra_coef_czb_full_off = 1.0            ! i.e. always on MS
    
    xtra_coef_a_l_nb_czb = 0.1
    xtra_dist_a_l_nb_czb = 10.0
    
    xtra_coef_b_l_nb_czb = 0.1
    xtra_dist_b_l_nb_czb = 3.0
    
    okay_to_remesh = .true.

!!! Input and Output
    max_num_profile_models = 500
    profile_interval = 100
    history_interval = 1
    terminal_cnt = 10
    write_header_frequency = 5
    
  ! GYRE output
    write_pulse_info_with_profile = .true.
    pulse_info_format = 'GYRE'
    add_atmosphere_to_pulse_info = .true.

/ ! end of controls namelist
\end{verbatim}

\end{appendix}

\end{document}